\documentclass{article}
\pdfoutput=1

\usepackage[english]{babel}
\usepackage{siunitx}
\usepackage{amsmath}
\usepackage{nicefrac}

\usepackage{tikz}
\usepackage{varwidth}
\usepackage{amssymb}
\usepackage{authblk}

\usepackage{hyperref}
\usepackage{listings}

\definecolor{codegreen}{rgb}{0,0.6,0}
\definecolor{codegray}{rgb}{0.5,0.5,0.5}
\definecolor{codepurple}{rgb}{0.58,0,0.82}
\definecolor{backcolour}{rgb}{0.95,0.95,0.92}

\lstdefinestyle{python_custom}{
    commentstyle=\color{codegreen},
    keywordstyle=\color{magenta},
    numberstyle=\tiny\color{codegray},
    stringstyle=\color{codepurple},
    basicstyle=\ttfamily\scriptsize,
    breakatwhitespace=false,
    breaklines=true,
    captionpos=b,
    keepspaces=true,
    numbers=left,
    numbersep=5pt,
    showspaces=false,
    showstringspaces=false,
    showtabs=false,
    tabsize=2
}
\DeclareSIUnit\pixel{px}
\makeatletter

\newcommand{\imag}{\mathrm{\mathbf{i}}}
\newcommand{\kv}{\mathrm{k}_0}

\begin{document}                  
\date{}

\title{The Scatman: an approximate method for fast wide-angle scattering simulations}

\author[1]{Alessandro Colombo\thanks{alcolombo@phys.ethz.ch}}
\author[1]{Julian Zimmermann}
\author[2]{Bruno Langbehn}
\author[2]{Thomas M{\"o}ller}
\author[3]{Christian Peltz}
\author[3]{Katharina Sander}
\author[3]{Björn Kruse}
\author[3]{Paul Tümmler}
\author[3]{Ingo Barke}
\author[1]{Daniela Rupp\thanks{daniela.rupp@phys.ethz.ch}}
\author[3]{Thomas Fennel\thanks{thomas.fennel@uni-rostock.de}}

\affil[1]{Laboratory for Solid State Physics, ETH Z{\"u}rich, 8093 Z{\"u}rich, Switzerland}
\affil[2]{Institute for Optics and Atomic Physics, Technical University Berlin, 10623 Berlin, Germany}
\affil[3]{Department of Physics, University of Rostock, 18051 Rostock, Germany}

\maketitle

\begin{abstract}
Single-shot Coherent Diffraction Imaging (CDI) is a powerful approach to characterize the structure and dynamics of isolated nanoscale objects such as single viruses, aerosols, nanocrystals or droplets.
Using X-ray wavelengths, the diffraction images in CDI experiments usually cover only small scattering angles of few degrees. These small-angle patterns represent the magnitude of the Fourier transform of the two-dimensional projection of the sample’s electron density, which can be reconstructed efficiently but lacks any depth information.
In cases where the diffracted signal can be measured up to scattering angles exceeding $\sim 10^\circ$, i.e. in the wide-angle regime, three-dimensional morphological information of the target is contained in a single-shot diffraction pattern. However, the extraction of the 3D structural information is no longer straightforward and defines the key challenge in wide-angle CDI. So far, the most convenient approach relies on iterative forward fitting of the scattering pattern using scattering simulations. Here we present the Scatman, an approximate and fast numerical tool for the simulation and iterative fitting of wide-angle scattering images of isolated samples.  Furthermore, we publish and describe in detail our Open Source software implementation of the Scatman algorithm, PyScatman. The Scatman approach, which was already applied in previous works for forward-fitting-based shape retrieval, adopts the Multi-Slice Fourier Transform method. The effects of optical properties are partially included, yielding quantitative results for weakly scattering samples. PyScatman is capable of computing wide-angle scattering patterns in few milliseconds even on consumer-level computing hardware. The high computational efficiency of PyScatman enables effective data analysis based on model fitting, thus representing an important step towards a systematic application of 3D Coherent Diffraction Imaging from single wide-angle diffraction patterns in various scientific communities.

\end{abstract}
\section{Introduction}

Coherent Diffraction Imaging (CDI) aims at retrieving an isolated sample's spatial information from the far-field amplitude of a highly coherent and monochromatic light beam that has scattered off the sample \cite{chapman2010coherent, miao2011coherent, seibert2011single}. The great advantage of CDI is its \emph{lensless} setup, making it suitable for those wavelength regions where lenses are hard or even impossible to manufacture. Thus, the spatial resolution in CDI is, in principle, only dependent on the radiation wavelength and on the maximum scattering angle at which the scattering signal can be recorded on a detector.

For small-angle scattering (SAS) conditions \cite{guinier1955small}, 
and assuming first Born's approximation \cite{born1926quantenmechanik}, the 2D scattering image can be efficiently computed by calculating the squared absolute value of a Fourier Transform (FT) of the imaged sample's 2D electron density projection.
This relationship between the sample and the diffraction within the SAS regime is at the basis of the \emph{original} CDI approach, experimentally demonstrated for the first time in 1999 \cite{miao1999extending}, where iterative \emph{phase retrieval} algorithms are employed to reconstruct the scattered field in the detector's plane in amplitude and phase\cite{fienup1982phase, marchesini2007invited}. Upon successful phase recovery, the real-space 2D projection of the sample can be directly computed\cite{loh2012fractal,seibert2011single,pedersoli2013mesoscale}.

Assuming that multiple diffraction patterns from equivalent objects are available, the SAS scheme has been shown to enable the reconstruction of the full 3D structure of a target.
The most obvious way is based on the \emph{tomographic} approach, where several diffraction patterns of the same sample are acquired at different orientations, giving a sufficient amount of 3D information in the reciprocal space to perform a 3D \emph{phase retrieval} process via suitable algorithms\cite{miao2006three, jiang2010quantitative, lundholm2018considerations, loh2010cryptotomography, loh2009reconstruction, ekeberg2015three}. 
In this respect, the recent advent of X-Ray Free Electron Laser (XFEL) sources
 \cite{feldhaus2005x,harmand2013achieving, barty2013molecular, chapman2006femtosecond} has opened new routes for characterizing so far elusive objects, thanks to their ultra-short and ultra-high intensity pulses, enabling to record meaningful scattering signal before the object is destroyed, a scheme that has therefore been termed \emph{diffraction before destruction} \cite{Chapman2014}. As a result, however, each sample can only provide a single diffraction pattern before being destroyed by the laser radiation. Thus, the 3D \emph{tomographic} approach is viable only if many replicas of the same sample are available \cite{ekeberg2015three}. Although additional shape information or symmetry constraints on the sample can in principle allow for a shape retrieval from a single SAS diffraction image \cite{xu2014single}, a full 3D reconstruction of non-replicable samples with unconstrained shapes is impossible to perform with SAS experiments. 
The requirement for additional constraints for reconstructing 3D information from an SAS experiment is a result of the fact that the magnitude of the maximum transfer momentum $\vec{q}$ acquired by the scattering detector is much smaller than the radiation momentum  $\kv$. Thus, as intuitively presented in \cite{barke20153d}, the acquired transfer momenta lie essentially in the plane orthogonal to the beam propagation direction  (see also Fig. \ref{fig:momentum}b in Sec. \ref{sec:scatman_routine}), and the sample's depth information is, in practice, completely lost.

The limitation to 2D-only information can be overcome in the so-called wide-angle scattering regime (WAS). Most importantly, in this regime the 2D diffraction patterns contain 3D information, because of comparable magnitudes of the transfer momentum $\vec{q}$ and the wave vector $\kv$ (see Fig. \ref{fig:momentum}a in Sec. \ref{sec:scatman_routine}).
As shown in  \cite{barke20153d}, in this scenario different parts of the scattering pattern carry details about different 2D projections of the density - establishing the possibility to extract tomographic information from a single image.
The primary shortcoming of experiments in the WAS regime is that the scattering patterns cannot be converted into shape information in such a straightforward way as in the SAS regime, where the field represents the 2D Fourier transform of the density projection. Some attempts to numerically invert single WAS patterns were made \cite{raines2010three}: however, the stability and reliability of such approaches are still debated within the community \cite{wang2011non,miao2011potential}.

Therefore, the \emph{forward-fitting} approach, where a measured scattering pattern is compared to scattering simulations for appropriately parameterized sample's shapes, is currently the most general and practicable  approach to invert CDI data taken under WAS conditions. 
To perform such a \emph{forward-fitting} analysis, a model that describes the sample's morphology depending on a set of free parameters has to be selected. Then, those parameters' values are varied using stochastic and/or deterministic optimization algorithms to minimize the discrepancy between the experimental diffraction data and the scattering simulation. In this procedure, the simulation of scattering patterns is the most challenging and computationally expensive task, highlighting the urgent need for fast forward-simulation approaches.

If the simulation runtime is uncritical, e.g. for benchmarking purposes, or for cases with high symmetry, several approaches are available that enable to compute the exact solution to the scattering problem.
The first method is based on the analytical solution for sufficiently simple geometries, such as the Mie solution to the Maxwell equations \cite{hahn2009light}, with which the scattered far-field can be calculated as a series expansion into vector wave harmonics up to arbitrary accuracy. However, such analytically motivated treatment is only applicable to simple sample shapes, like a sphere \cite{bohren2008absorption} or a coated sphere \cite{aden1951scattering}.
A second option is to compute the scattering by solving Maxwell's equations numerically, e.g. via the Finite-Difference Time-Domain (FDTD) method \cite{taflove1980application, varin2012attosecond} or using Green's function based approaches such as the Discrete Dipole Approximation (DDA)  \cite{purcell1973scattering, sander2015influence}. These numerical methods allow simulations of light-matter interaction with no restrictions on the sample's shape. However, FDTD or DDA calculations are computationally cost-intensive, as the whole computational domain has to be represented on a grid at a sufficiently fine scale, for which the temporal evolution (FDTD case) or the iterative solution for the fields evolution (DDA case) have to be calculated. The demanding computational conditions render the methods aiming at the unrestricted full solution of Maxwell's equations  impractical for the use-case of simulating more than a few diffraction images. 
Therefore, suitable approximate methods are highly attractive for data analysis of wide-angle CDI. In this paper we present a fast, flexible, and intuitive approximate simulation suite: The Scatman.

\begin{figure}
  \centering
  \includegraphics[width=\textwidth]{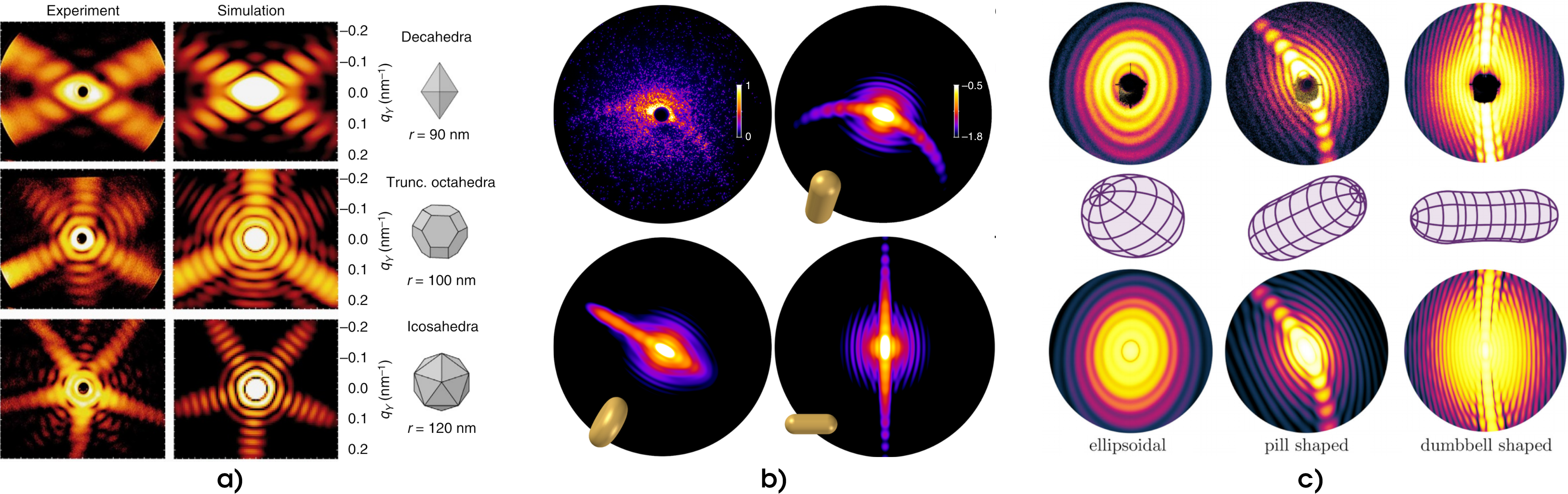}
  \caption{Examples extracted from previous works that made use of the Scatman approach for wide-angle scattering data analysis. In a), adapted from \cite{barke20153d}, soft X-Rays are used to study the 3D structure of silver nano-particles, by comparing the experimental data with the simulation. In b), adapted from \cite{rupp2017coherent}, comparisons between experimental data and simulations demonstrated the feasibility of Coherent Diffraction Imaging with High Harmonic Generation sources. In c), adapted from \cite{langbehn2018three}, a fitting between the Scatman result and experimental diffraction patterns revealed the 3D shapes of superfluid helium nanodroplets. }
  \label{fig:papers}
\end{figure}

The Scatman's core was originally conceived by the authors of Barke et al. \cite{barke20153d}, and computes wide-angle coherent scattering images of isolated samples. It has already proven successful for data analysis of WAS experiments \cite{barke20153d, rupp2017coherent,langbehn2018three,zimmermann2019deep}: examples ranging from silver nanocrystals to spinning superfluid helium droplets are depicted in Fig. \ref{fig:papers}.
A continued development led to the refined, generalized, and concise form of the Scatman presented here.

The two following sections of this paper are dedicated to the analytical framework and motivation of the approach, based on the Multi-Slice Fourier Transform (MSFT) technique \cite{cowley1957scattering,self1983practical,reinhard1997size, hare1994near,barke20153d}, and its translation into a numerical form. Section 4 focuses on the comparison between the simulation results of the Scatman and exact, analytical calculations based on Mie theory for a spherical sample. It provides intuition about the region of applicability of the Scatman, whose results can be quantitatively close or just qualitatively usable, depending on the sample's properties.
The final sections, 5 and 6, present our Scatman reference implementation, called PyScatman, published along with this paper as Open Source software. PyScatman is released as a Python module that provides an easy interface to the user, and incorporates \emph{state of the art} programming techniques to yield a high computational efficiency.

\section{The Scatman routine} \label{sec:scatman_routine}

\begin{figure}
  \includegraphics[width=\textwidth]{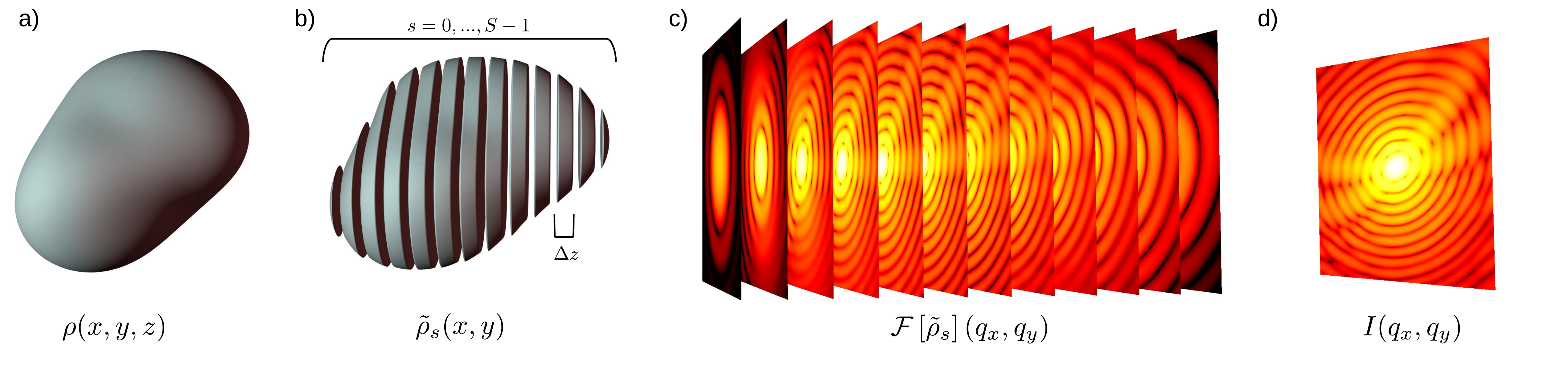}
  \caption{Schematic representation of the Scatman's MSFT approach. In a), the sample as a whole is defined by its \emph{scattering strength} $\rho$, which depends on the spatial distribution of the complex refractive index $n$. In b), the sample is split into $S$ slices, where, for each slice $s$, the scattering density $\tilde{\rho}_s$ is determined by the slice's optical properties. In c), the scattered far field is computed for each slice $s$. In the last step d), the scattering of the slices is summed up with a proper phase correction and subsequently squared to simulate the recorded diffraction pattern on the detector. Please note that, for presentation purposes, c) only shows the scattering signal's squared amplitude for every slice, while the actual scattered wavefield is still a complex function at this point.}
  \label{fig:msft}
\end{figure}

The Scatman is based on the MSFT approach, originally developed for electron scattering \cite{cowley1957scattering,self1983practical,reinhard1997size}. The MSFT routine was already applied  in X-ray diffraction experiments for fixed targets \cite{hare1994near}, as well as for recovering the topology of individual silver and helium nanoparticles in free flight \cite{barke20153d, langbehn2018three} . 
A schematic overview on the MSFT method is shown in Fig. \ref{fig:msft}. Roughly speaking, the simulation is based on the partitioning of the spatial domain into slices (Fig. \ref{fig:msft}a and \ref{fig:msft}b). The scattering contribution from each slice is computed independently via a Fourier Transform operation (Fig. \ref{fig:msft}c) and then summed up with a proper phase correction to compose the final scattering pattern (Fig. \ref{fig:msft}d).
This section briefly revisits the mathematical derivation of the approach, particularly focusing on how the effects of the sample's refractive index are effectively incorporated into the Scatman's simulation.

\begin{figure} 
  \centering
  \includegraphics[width=0.75\textwidth]{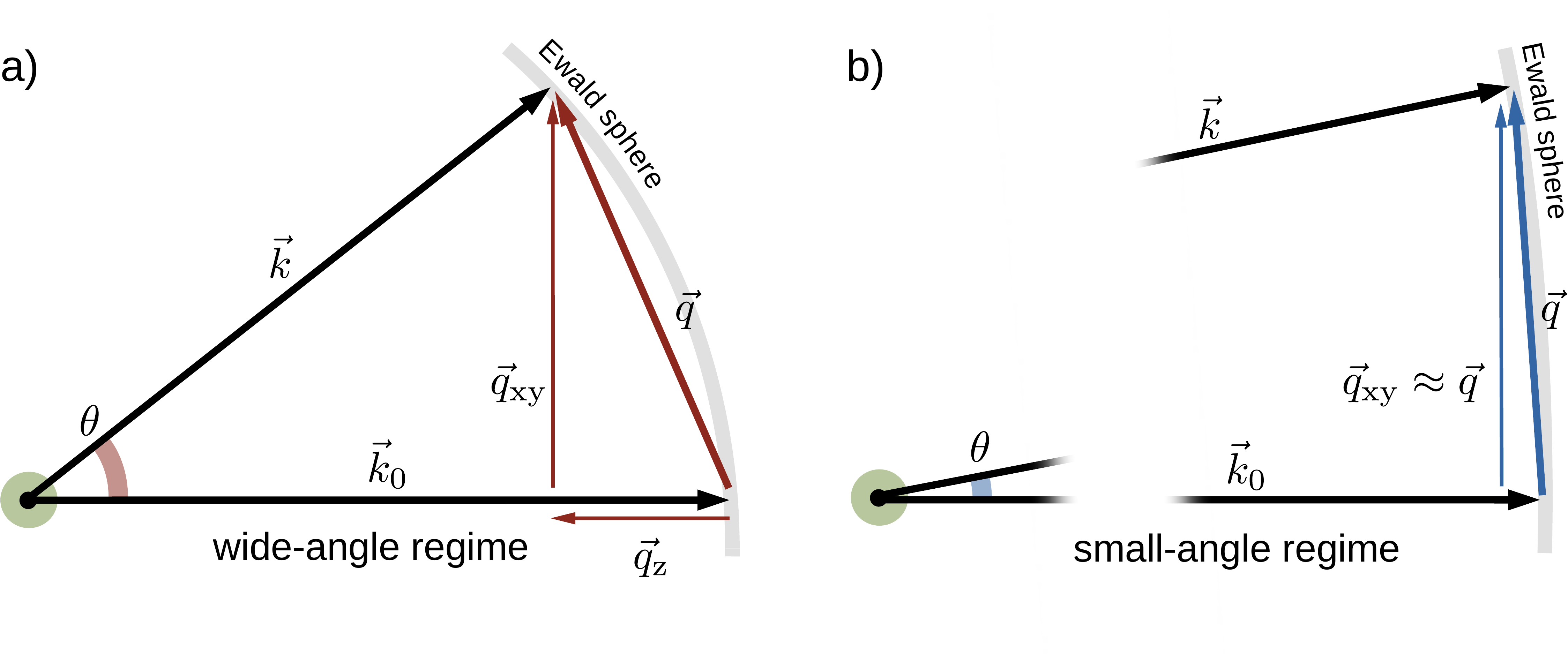}
  \caption{Schematic view on wide-angle and small-angle regimes and on the notation used in this paper for the momentum vectors. The transfer momentum $\vec{q}$ is defined as the difference between the incoming wave-vector $\vec{\kv}$ and the scattering vector $\vec{k}$. The transfer momentum $\vec{q}$ can be decomposed into its axial component $\vec{q_z}$, i.e. the component parallel to the incoming radiation assumed to travel along the $z$ axis, and $\vec{q}_{xy}$. The \emph{wide-angle} regime is depicted in a), where the axial component of the transfer momentum $\vec{q_z}$ is non-negligible thanks to the large scattering angle $\theta$. For comparison, b) shows the same scheme in the \emph{small-angle} regime, where the scattering angle $\theta$ is sufficiently small to neglect the axial component of $\vec{q}$. In both a) and b) the \emph{Ewald sphere} is shown in gray. }
  \label{fig:momentum}
\end{figure}

For deriving the method, we start from the well-known Born's approximation \cite{born1926quantenmechanik}, which defines the scattered field  $\Psi(\vec{q})$ in the far-field condition as:
\begin{equation}\label{eq:Born_equation}
    \Psi (\vec{q}) \propto \int \! \mathrm{d}\vec{r} \, \rho(\vec{r}) \, e^{ \imag \vec{q}\cdot \vec{r}} 
\end{equation}
where $\rho(\vec{r})$ defines the scattering \emph{strength} in space and $\vec{q}$ is the transfer momentum, schematically shown in Fig. \ref{fig:momentum}. 
The integral in Eq. \eqref{eq:Born_equation}, which is in practice a 3D Fourier Transform of the scattering strength $\rho(\vec{r})$, can be re-written in the following form:

\begin{equation}\label{eq:MSFT_equation}
  \begin{split}
    \Psi (\vec{q}) & \propto \iiint \! \mathrm{d}x \, \mathrm{d}y \, \mathrm{d}z \, \rho(x,y,z) \, \, e^{ \imag [x q_x + y q_y + z q_z]} \\
    & \sim \sum_s e^{ \imag q_z s \Delta z } \iint \! \mathrm{d}x \, \mathrm{d}y \, \left [ \int_{s\Delta z}^{s\Delta z + \Delta z} \mathrm{d}z \, \rho(x,y,z) \right ] \, \, e^{ \imag [x q_x + y q_y]} \\ 
    & = \sum_s e^{ \imag q_z s \Delta z }  \mathcal{F} \left [ \tilde{\rho}_s(x,y) \right ] (q_x, q_y)
   \end{split} .
\end{equation}

The first step in Eq. \eqref{eq:MSFT_equation} is the explicit formulation of Eq. \eqref{eq:Born_equation} in cartesian coordinates, where the coordinate system is chosen such that the $z$ axis is parallel to the beam propagation direction: from now on, this axis will also be addressed as the \emph{axial} direction. The second step of the equation is the approximation of the integral along $z$ with a discrete summation over the slices with integer index $s$: such approximation holds given the condition that $\Delta z \ll \frac{\pi}{q_z}$. As depicted in Fig. \ref{fig:msft}b, this operation represents a partitioning of the spatial domain into slices of size $\Delta z$ along the axial direction, being the core of the MSFT approach. 
 In the last step, the integral over the $x$ and $y$ direction is rewritten as a two-dimensional Fourier Transform. Moreover, the integral of the scattering strength $\rho$ over the slice $s$ in the axial direction is defined as $\tilde{\rho}_s$. 
 
As long as only monochromatic radiation with momentum $\kv$ is considered and the scattering event is assumed to be completely \emph{elastic}, it is convenient to re-write the axial component of the transfer momentum $q_z$ as function of $q_x$ and $q_y$:
 \begin{equation}\label{eq:qz}
 \begin{split}
    q_z(q_x, q_y) &= \kv \left[1-\cos(\theta(q_x,q_y)) \right]  \\
    \text{with} \,\,\,\, \theta(q_x,q_y) &= \arccos( \sqrt{1 - \frac{q_x^2 + q_y^2}{\kv^2}})
  \end{split}
\end{equation}
where $\theta$ is the scattering angle. Eq. \eqref{eq:qz}, which can be intuitively derived from Fig. \ref{fig:momentum}a through geometrical considerations, enables to rewrite Eq. \eqref{eq:MSFT_equation} as:
 \begin{equation}\label{eq:MSFT_equation_2}
    \Psi(q_x, q_y) = \sum_s e^{ \imag \kv \left[1-\cos(\theta(q_x,q_y)) \right] s \Delta z }  \mathcal{F} \left [ \tilde{\rho}_s(x,y) \right ] (q_x, q_y) \, . 
\end{equation} 
The scattered field $\Psi (q_x, q_y)$ in Eq. \eqref{eq:MSFT_equation_2}  is the sum of the scattering contributions from all $s$ slices, with a proper phase factor that depends on the scattering angle and on the slice's position on the $z$ axis.
 
We continue the derivation of the method by defining the scattering strength $\rho(\vec{r})$. In particular, at high photon energies, the strength of the scattering is related to the amount of electronic charges that contribute to the scattering. These are described by the dielectric polarization density $P(\vec{r})$, defined as:
\begin{equation} \label{eq:scattering_strength}
P(\vec{r}) = \varepsilon_0 \chi_e (\vec{r}) E(\vec{r}) \approx  \varepsilon_0 \left [ n^2 (\vec{r}) -1 \right] E(\vec{r})
\end{equation}
where $E(\vec{r})$ is the electric field and $\chi_e(\vec{r})$ is the electric susceptibility. The latter has been then rewritten as function of the complex refractive index $n$, exploiting its relationship with the relative permittivity $ \varepsilon_r = \chi_e + 1$ , which is equivalent to the squared refractive index for non-magnetic materials, i.e. $n^2 = \varepsilon_r \mu_r \approx  \varepsilon_r$.

The Born's approximation in Eq. \eqref{eq:Born_equation} assumes that the incoming electric field is not affected by the presence of the sample, and considers a planar wave with constant amplitude, momentum and phase along the full path. In this view, the scattering strength defined in Eq. \eqref{eq:Born_equation} is proportional to the electric susceptibility, i.e.  $\rho(\vec{r}) \propto \left [ n^2 (\vec{r}) -1 \right] $, allowing the definition of the scattering strength for a slice in Eq. \eqref{eq:MSFT_equation_2} as $\tilde{\rho}_s(x,y) \propto \left [ \tilde{n}_s^2 (x,y) -1 \right] $, where $\tilde{n}_s$ are the optical properties averaged over the slice thickness.
However, especially when considering wide-angle scattering, the refractive index of the sample will modify the field. It is possible to partially take this effect into account by defining the scattering strength  for a given slice $s$ in the following form:
\begin{equation} \label{eq:scattering_strength_slice}
\tilde{\rho}_s(x,y) \propto \left [ \tilde{n}_s^2 (x,y) -1 \right] \frac{E_s(x,y)}{E_0 \, e^{\imag \kv  s  \Delta z }} 
\end{equation}
where $E_s(x,y)$ is the field actually impinging on slice $s$ of the sample, while $E_0 \, e^{\imag \kv  s  \Delta z }$ is the field as it would travel unaffected by the presence of the sample (the field taken into account by the Born's approximation in Eq. \eqref{eq:Born_equation}). The ratio between the two fields can be interpreted as a \emph{correction} applied to the unmodified field assumed in the first Born's approximation, making $\tilde{\rho}_s(x,y)$ in Eq. \eqref{eq:scattering_strength_slice} an \emph{effective} scattering strength. This correction factor allows to approximately include the effects of the sample's optical properties on the electric field incoming to the slice, while the scattered field is still the unaffected one considered in the Born's approximation.

For a more intuitive presentation of how the incoming field $E_s (x,y)$  impinging on the slice $s$ is treated in the Scatman approach, it is now convenient to rewrite the sample's refractive index $n$ in the following form:
\begin{equation} \label{eq:refractive_index}
  n(\vec{r})=1-\delta(\vec{r}) + \imag \beta(\vec{r}).
\end{equation}
Here, $\delta$ defines the deviation of the real part of $n$ from unity, and is responsible for the change of the light's phase velocity in the sample, causing also the phenomena of \emph{refraction} and \emph{reflection}. On the other side, $\beta$, often called \emph{extinction} or \emph{absorption} coefficient, defines how much the radiation is damped when traveling in the sample \cite{lambert1760photometria, beer1852bestimmung}. This notation for the refractive index is convenient in the X-Ray regime, where $|n|$ is very close to unity, and will be extensively used in this manuscript.

An exact description of how the field distribution in the sample is affected by $\delta$ and $\beta$, regarding the field's amplitude, phase, and propagation direction, is highly demanding and essentially requires again the full solution of the scattering problem. However, in the limit of sufficiently small $\delta$ and $\beta$, it is possible to assume the so-called \emph{projection approximation} \cite{paganin2006coherent}, reducing the expression for the propagation of the electric field $E_s$ at slice $s$ to the following form:
\begin{equation} \label{eq:E_propagation}
\begin{split}
E_s(x,y) \approx & \, E_{s-1}(x,y) \, e^{\imag \, \mathrm{\kv} \, \Delta z \, \tilde{n}_{s-1} (x,y) } \\
 = & \, E_{s-1}(x,y) \, e^{\imag \, \mathrm{\kv} \, \Delta z} \, e^{- \imag \, \mathrm{\kv} \, \Delta z \, \tilde{\delta}_{s-1} (x,y)} \, e^{- \mathrm{\kv} \, \Delta z \, \tilde{\beta}_{s-1} (x,y)}
\end{split}
\end{equation}
where $\tilde{\delta}$ and $\tilde{\beta}$ are the values of $\delta$ and $\beta$ averaged over the slice thickness $\Delta z$, and $\mathrm{\kv} $ is the radiation wavenumber. This approximation locally assumes an axial propagation through a homogenous medium. Eq. \eqref{eq:E_propagation} recursively describes how the field impinging on slice $s$ is modified by taking into account the effects of all the preceding slices. A first strong assumption made by Eq. \eqref{eq:E_propagation} is that $\delta$ is sufficiently small to neglect changes in the field propagation direction due to \emph{refractions} and \emph{reflections}, i.e. the electric field always propagates in the axial direction, even within the sample. Moreover, $\delta$ and $\beta$ are assumed to be sufficiently small to neglect their influence on the radiation scattered by the preceding slices, i.e. \emph{secondary} scattering is completely neglected (for the discussion of the resulting limitations, see Section 4). In practice, at a given slice $s$, $\tilde{\delta}_s$ introduces a \emph{phase shift} in the field, while $\tilde{\beta}_s$ exponentially dampens the field magnitude.

Finally, the scattering strength (Eq. \eqref{eq:scattering_strength_slice}) has to be inserted into Eq. \eqref{eq:MSFT_equation_2}. The electric field in the denominator of Eq. \eqref{eq:scattering_strength_slice}, which is independent on $q_x$ and $q_y$, can be pulled out of the Fourier Transform and simplifies the global phase pre-factor of the slice. This operation, combined with the formula for the approximated field propagation in Eq. \eqref{eq:E_propagation}, yields the \emph{main equation} of the Scatman approach:
\begin{equation}\label{eq:MSFT_master}
  \begin{split}
     \Psi &  (q_x, q_y)  \! \propto \!  \sum_s e^{- \imag\, \sqrt{\kv^2 - q_x^2 - q_y^2} \, s \, \Delta z} \, \mathcal{F} \left[ \left (\tilde{n}_s^2  -1 \right ) E_s \right ] (q_x, q_y) \\
    & \text{with } \,\,\,\,\,\,
    \begin{split}
     & E_s(x,y)   = E_{s-1}(x,y) \, e^{\imag \, \mathrm{\kv} \, \Delta z \, \tilde{n}_{s-1} (x,y)}\\
     & E_{s=0}(x,y) = E_0 \,\,\, .\\
    \end{split}
  \end{split}
\end{equation}
Here the scattered field is defined based on the spatial distribution of the sample's optical properties.
Although Eq. \eqref{eq:MSFT_master} predicts the scattered electric field, only its squared amplitude $I(q_x, q_y) = \left| \Psi (q_x, q_y)  \right|^2$ is physically measured in CDI experiments \cite{gaffney2007imaging, chapman2010coherent, seibert2011single, ekeberg2015three} and should be taken into account for actual simulations.

\section{Numerical implementation}

\begin{figure}
  \centering 
  \includegraphics[width=\textwidth]{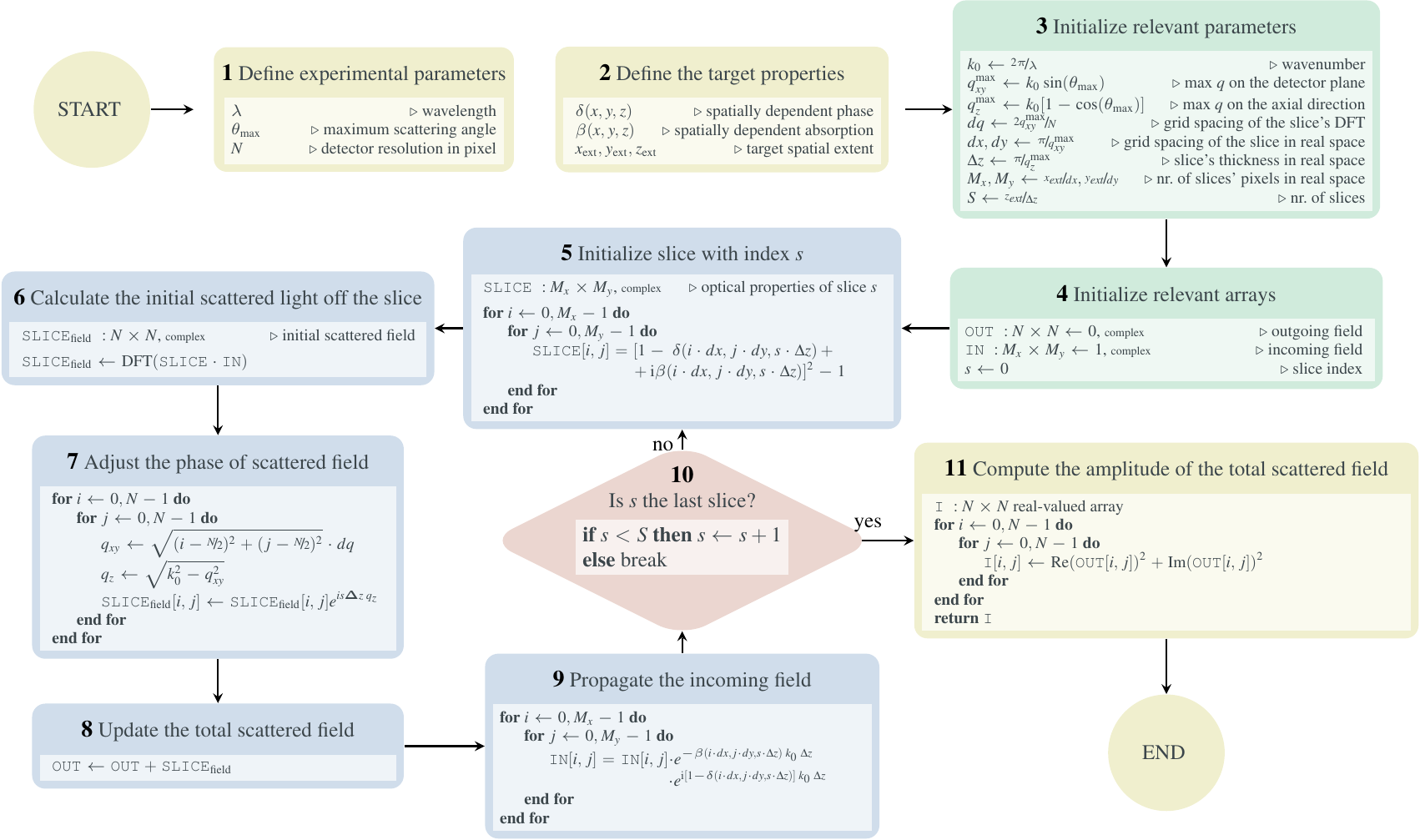}
  \caption{A flowchart of the conceptual core of the Scatman in numerical form is shown. Yellow blocks indicate I/O operations. Green blocks contain data preparation. The main loop of the program, which carries out the majority of all calculations, is highlighted in blue. Each block contains a pseudo-code schematically showing the numerical calculation. The abbreviation DFT stands for Discrete Fourier Transform, practically computed through the Fast Fourier Transform algorithm \cite{cooley1965algorithm}.}
  \label{fig:flowchart}
\end{figure}

In this section the concrete numerical implementation of the Scatman is provided.
A flowchart of the program  is shown in Fig. \ref{fig:flowchart}, and its blocks are described in the following paragraphs.

\subsection{Setting up the \emph{virtual} experiment}

\paragraph{Read-in the user-defined parameters (Block 1 + 2)}

At the beginning of the program, the user-defined input parameters are read-in. These include experimental details (Block 1) like the irradiation wavelength $\lambda$, the maximal scattering angle $\theta_{max}$, and the detector resolution $N$ in pixels. $N=1000$ will result in a virtual detector of size $1000 \times 1000$ pixel. The virtual detector is centered on-axis in the z-direction, and every \emph{virtual} pixel has the same angular cross section.
Furthermore, the target is defined via a spatially dependent refractive index ($\delta$ and $\beta$) and the concrete dimensions of the target ($x_\text{ext}$, $y_\text{ext}$, $z_\text{ext}$), as listed in Block 2.
With this set of parameters the experimental setup is uniquely defined.

\paragraph{Initialization (Block 3 + 4)}

Before entering the program's main loop, additional parameters are derived from the user-defined parameters, and the relevant arrays are initialized.

Here, the maximal components of the scattering vector on the $xy$ plane (also called \emph{detector} plane) and in the axial direction, $q_{xy}^\text{max}$ and $q_{z}^\text{max}$, are calculated. As the Fourier transform in Eq. \eqref{eq:MSFT_master} is numerically computed in the discrete form, $q_{xy}^\text{max}$ is necessary to assign a corresponding size $dx$ and $dy$ to the slices' pixels in real space, and consequently its spatial extension in pixels, $M_x$ and $M_y$. The same applies to the axial direction, where $q_z$ defines the slice thickness $\Delta z$, and thus the total amount of slices $S$.

Two complex numerical arrays are then initialized to hold the 2D wavefronts of the outgoing and the incoming wavefield (Block 4). While the outgoing wavefield is initialized to zero, the incoming field is initialized to 1, which is equivalent to a spatially coherent and plane wavefront. The desired diffraction image can now be iteratively computed within the program's main loop.

\subsection{The main loop}

The main loop is the core of the program. Every iteration within it calculates the scattering contribution of one slice of the sample. Each slice's input field is the original plane wave shaped by the optical properties of the sample up to the slice of interest. The scattered radiation is then corrected with a proper phase factor.

\paragraph{Calculating the local scattered field (Block 5 + 6 + 7)}

The first step required to compute the slice's scattered field is to render the slice's scattering potential (Block 5) through the computation of the sample's optical properties at the proper spatial coordinates.
Calculating the slice's scattering contribution is, then, the subject of Blocks 6 and 7, where, for building intuition, both multiplicative terms are treated in their own blocks. First, the Fourier transform of the product between the incoming field and the slice yields the far-field scattering contribution of the latter (Block 6). Then, the proper phase correction is applied (Block 7). It should be noted that the scattering vector's components are derived in pixel units from their off-axis distance relative to the z-axis. The final wavefield is then stored in the complex array $\mathtt{SLICE}_\text{field}$.

\paragraph{Updating the total scattered field and computing the incident field for the next slice (Block 8 + 9)}
The total scattered field is updated in block 8 by adding the scattering contribution of slice $s$ to the ones of all the previous slices.

Then, in block 9, the incoming field for the next slice is prepared by propagating the field through slice $s$ along the z-axis, following the definition in Eq. \eqref{eq:E_propagation}. 
The decoupling of the total scattered field (Block 8) and the incoming field for any subsequent slice (Block 9) is enforcing a central assumption within the Scatman: multiple sequential scattering events are not allowed to occur.
Every slice is, thereby, irradiated by a plane wave, which traveled through the medium without scattering up to the point where it encounters the slice.

\subsection{Preparing the output of the Scatman}

After the main loop iterates over all slices in the virtual medium, i.e. when block 10 reaches the loop break condition, the Scatman's final piece of code prepares the output to match the experimental conditions. In the simplest case, this is just computing the absolute squared value of all slices' total scattered field (Block 11). However, it can include the modeling of detector artifacts, straylight during the experiment, or any other experimental effect desired to model onto the simulated scattering images.

\section{Evaluation of the Scatman using exact simulations}
\label{sec:comp_mie_solution}

As highlighted in the introduction, the Scatman program is an alternative to the computationally intensive, but versatile, numerical simulations such as FDTD or DDA methods, and to the fast, but topologically restrictive, Mie's analytical solutions to Maxwell equations. However, as underlined in Section \ref{sec:scatman_routine}, the Scatman's capability of being both fast and versatile is traded off against the accuracy of the simulation results, which heavily depend on the choice of the simulation parameters.

During the mathematical formulation of the approach in Section \ref{sec:scatman_routine}, some approximations were involved. Most of them imply the assumption that the optical properties of the sample of interest only slightly differ from the ones of the surrounding medium (assumed here to be vacuum). Therefore, the Scatman approach is unable to quantitatively reproduce features of the scattering images when relatively large variations of the refractive index are present, e.g. close to electronic resonances. 
Still, probing how \emph{small} the variation of the refractive index has to be is of great interest to the user, to decide whether it is preferable to rely on alternative and more accurate methods for data analysis. 
This section provides an overview of the capabilities and limitations of the Scatman program, where we compare the simulation results with analytical diffraction patterns obtained via Mie theory.

\begin{figure}
  \includegraphics[width=\textwidth]{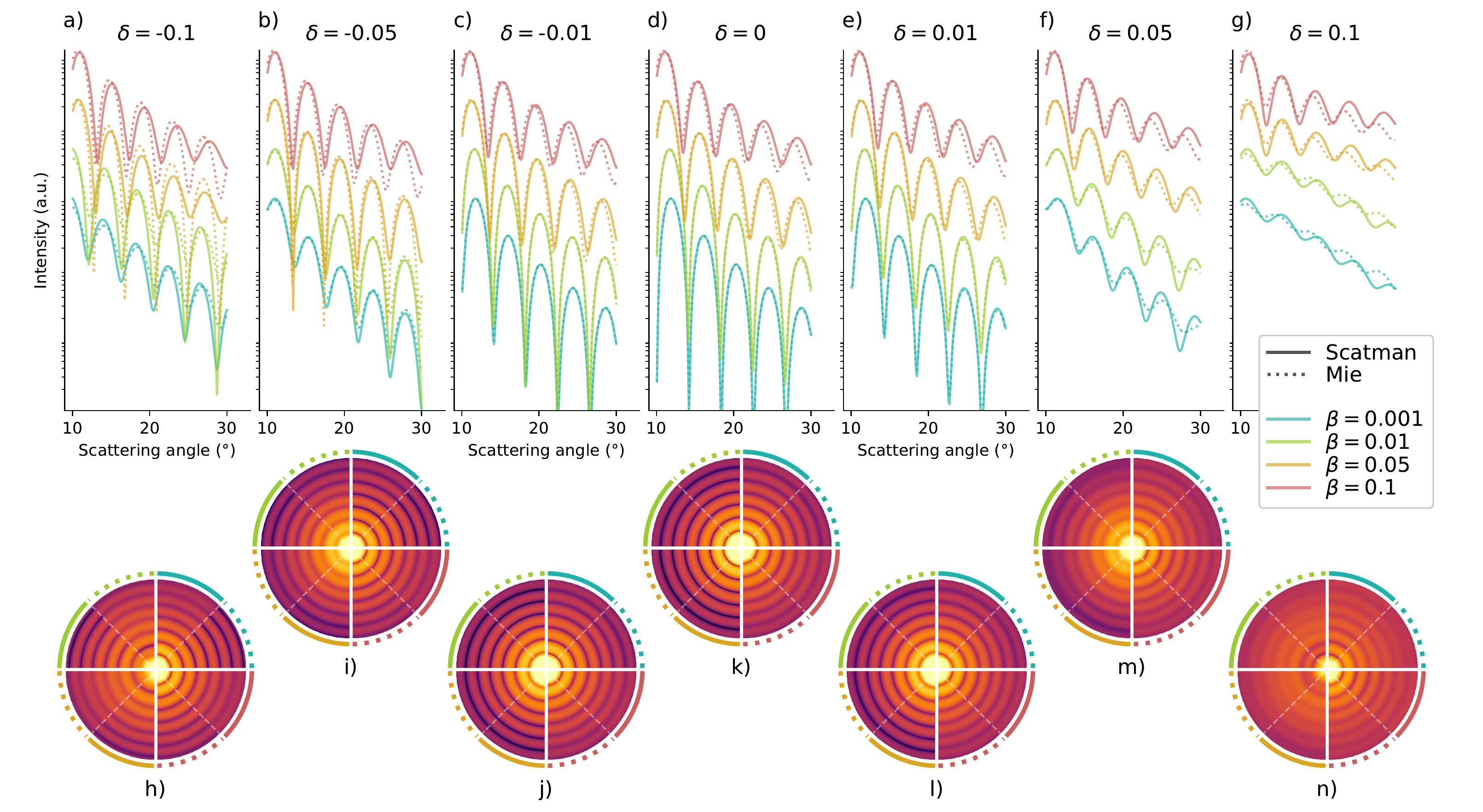}
  \caption{Comparison of radial profiles between the Scatman and Mie calculations. The figure is split into two rows that share a common legend, which is placed in between both rows. In the upper part, from a) to g), 28 radial profiles are shown that correspond to 28 combinations of $\delta$ and $\beta$ for a fixed spherical target of radius 7$\lambda$. For each combination of $\delta$ and $\beta$ the Scatman approximation along the exact Mie results are plotted in solid and dashed lines, respectively.
    In the lower part, the seven subplots from the top part are translated into seven diffraction images in h) to n), where the intensity of the scattering signal is encoded in logarithmic color scale. Every diffraction image is partitioned into eight segments. These eight segments correspond to the eight line-plots shown in the associated plot from the top row. Every $\delta$ / $\beta$ pair combination takes up a quarter of every diffraction image, where the solid and dashed lines surrounding the diffraction image indicate that the quarter is showing either the approximation of the Scatman program (solid line) or the exact Mie solution (dashed line).
  }
  \label{fig:mie_comp}
\end{figure}

Fig. \ref{fig:mie_comp} shows the results for 28 scattering simulations for a spherical target, with a different pair of $\delta$ and $\beta$ values each. The figure is split into two rows that show the radial profiles and the diffraction patterns, respectively. Both rows share a common legend, which is placed in between: solid lines represent the Scatman result, dashed lines the Mie solution, and the colors indicate different $\beta$ values. The top row shows seven subplots a) to g), where each subplot shows the scattering angle dependence of the scattered light from a spherical particle for a fixed value for $\delta$ and four values for $\beta$. The choice to limit the scattering angle to a range between \SIrange{10}{30}{\degree} corresponds to typical experimental scenarios for CDI experiments within the WAS regime \cite{rupp2017coherent, langbehn2018three, barke20153d}.
In every subplot and for every $\delta$ and $\beta$ pair, two calculations are shown, the solid line is the approximation of the Scatman program, and the dashed line is the exact Mie solution. The radius of the spherical target used for the simulations is fixed at 7$\lambda$, which enables to see the signature of different optical properties, and to  distinguish the maxima and minima of the interference as well. For a fair comparison between the two simulation methods, a normalization factor has to be defined: in this case, both Mie and Scatman profiles were normalized on their integral value computed between \SIrange{10}{30}{\degree}.

In the particular case of a spherical target, and assuming non-polarized light, the simulated diffraction image is identical in all scattering directions. This symmetry property is exploited in the bottom row of Fig. \ref{fig:mie_comp}. The seven subplots from the top row are translated into seven diffraction images h) to n), where every diffraction image is partitioned into eight segments. These eight segments correspond to the eight line-plots provided in the associated subplot from the top row. Every $\delta$ / $\beta$ pair combination takes up a quarter of every diffraction image, where the solid and dashed lines surrounding the diffraction image correspond to the approximation of the Scatman program (solid line) and the exact Mie solution (dashed line).

Therefore, the bottom row is not adding new data to the figure but is providing a strong proxy for building intuition on how the scattering image looks like. Furthermore, it enables a qualitative assertion on the diffraction images, which is often sufficient to deduce the sample's underlying topological properties. \cite{rupp2017coherent, langbehn2018three, barke20153d}.

The simplest case during this evaluation is for $\delta = 0$ in combination with the smallest $\beta$ value (\num{0.001}). There, the wavefield that propagates throughout the medium is identical in phase with a reference field propagating through the surrounding vacuum and only very weakly absorbed. The corresponding Scatman and Mie calculations are shown in blue in Fig. \ref{fig:mie_comp} d) and k).
The solid and dashed blue-indicated slices in the diffraction image in k) are indistinguishable by eye, just as the radial profiles in d). However, when increasing the absorption from \numrange{0.001}{0.01}, slight deviations become visible at high scattering angles, where the Scatman program produces a radial profile in which the maxima are shifted towards higher scattering angles, and the amplitude is slightly too high compared to the analytical results.

This behavior is core to all Scatman approximations where the absolute value of $\delta$ is comparably small ($| \delta | \lessapprox 0.1$). With increasing absorption, the Scatman overestimates the signal's total amplitude and shifts the extrema at larger scattering angles towards even larger scattering angles. Therefore, when $\delta$ is comparatively small, the quality of the Scatman's simulation is anticorrelated with the absorption in the medium.

The scenario strongly varies when larger values for $\delta$ are considered. There, the Scatman's behavior is more complicated, mostly due to the appearance of intricate resonance effects that arise from the interplay between the target's geometry and the wavefield. Such resonance effects are more pronounced for positive values for $\delta$ (refractive index smaller than unity), for example, observed in the atomic near-resonance regime, where the photon energy dependence of $\delta$ resembles a Fano profile. Thus, the assertion concerning the $\delta$-dependence must be split for positive and negative values. At negative values, broadly speaking in the off-resonance case, the deviations between the Scatman and the Mie simulation are mainly due to an overestimation of the amplitude with a relatively tiny shift of the extrema's positions in the radial profiles. However, at positive values for $\delta$, the deviations between both simulations are significant. With $\delta$ values above $0.1$, not shown here, the resulting radial profiles wildly differ from another.

Therefore, besides the anticorrelation with $\beta$ for small $\delta$, the second deduction that can be made here is that the Scatman produces worsening diffraction images with a more positive $\delta$ (refractive index smaller than unity). The pivotal point for this to happen is for $\delta \gtrapprox 0.1$.

So far, the comparison between the Scatman's results and Mie theory was restricted to a fixed target size. However, the features of scattering images of isolated nanoparticles vary significantly, also depending on the targets' size \cite{Mie1908, Bohren1998, Rupp2014}. Thus, a more exhaustive comparison, which includes also size effects, is presented in the Supplemental Material.

Concluding, the approximation employed by the Scatman program produces in most cases diffraction patterns of very high quality compared to the analytical Mie solution for spherical particles. In general, the quality of the routine is best when the phase term in the refractive index is small ($|\delta|\lessapprox0.02$). Then, only minor deviations are observed and the Scatman's approximation could even be used as a replacement for the Mie theory based solution. With increasing absolute values for $\delta$, the quality deteriorates as well, where larger positive values for $\delta$ are yielding worse results than larger negative values.

At low $\delta$ values, the absorption ($\beta$) is anti-correlated with the quality of the approximation, yielding high quality diffraction images when absorption is low. This relationship, however, is reversed for larger absolute values of $\delta$, where a larger amount of absorption yields a better comparison to the Mie theory calculations.

The next section now introduces PyScatman, a high-level Python front-end for the Scatman method.

\section{PyScatman: a high-level Python front-end}

In this section we present and explain the reference implementation of the Scatman in the form of a Python module, called PyScatman. The source code is available under the MIT license\footnote{\url{https://spdx.org/licenses/MIT.html}} at \url{https://gitlab.ethz.ch/nux/numerical-physics/pyscatman}, while the documentation can be found at \url{https://nux-group.gitlab.io/pyscatman/}.

The module is written in C++ \cite{c++2017iso} with bindings in Python using the PyBind11 C++ library \cite{jakob2019pybind11}. This hybrid approach enables us to maintain the highest possible simulation speed via compiled C++ code while keeping a Python-only user-friendly interface. The implementation is highly parallelized for multi-core CPUs, and takes advantage of Nvidia GPU accelerators via the CUDA library \cite{Nickolls2008}. At the current stage, PyScatman performs all the computations in \emph{single floating point precision} (32 bit).

In Section \ref{subsec:sim_example1} a fundamental example is provided and explained. There, an experiment is set up, an ideal detector is defined, and a simple shape is generated. 

Building on this, Section \ref{subsec:sim_example2} provides a more advanced example, where three shapes are generated using three different methods, and where a detector that simulates photon statistics is used. 
This second example is meant to highlight the great flexibility the PyScatman module offers in terms of defining a target's shape. 

Finally, in Section \ref{sec:performance}, the implementation is extensively benchmarked with respect to its execution time on either the CPU or the GPU using various shapes.

\subsection{A fundamental example}
\label{subsec:sim_example1}
In this section, a fundamental example is provided and explained. We demonstrate the basic functionality and show the easiest way for how to define the target's shape.

For discussing the elements in the script we will refer to the line numbers.
\begin{lstlisting}[language=Python, caption={A fundamental PyScatman example. Here, we set up the experiment, define a detector function and calculate the MSFT simulation for an ellipsoidal sample.}, label=script:pyscatman1]
import scatman

"""
1) Set up the experiment.
"""
scatman.set_experiment(
    wavelength=40,       # in nm
    angle=30,            # in degree
    resolution=1024      # in pixels
)

"""
2) Define the detector type.

MSFT is a virtual detector, returning the plain MSFT result
"""
detector = scatman.Detectors.MSFT()

"""
3) Define a shape.

Here, an ellipsoid model is used.
"""
shape_el = scatman.Shapes.Ellipsoid(
    a=100,               # the semi-axes in nm
    b=60,   
    c=60,
    delta=0.001,         # refractive index
    beta=0.01,
    latitude=40,         # the orientation in space
    longitude=30,        # see Fig. 6
    rotation=0
)

"""
4) Perform the MSFT calculation.

The detector class provides the simulation
result via its 'acquire' method.
"""
pattern_el = detector.acquire(
    shape=shape_el)
\end{lstlisting}

\paragraph{Define an experiment (line \numrange{1}{10}):} After the \verb+scatman+ module is imported, the experimental conditions are set up, by defining the irradiation wavelength in \si{\nano\metre}, the maximal scattering angle in degrees, and the desired detector resolution in \si{\pixel}. Within PyScatman, there is no preferred length unit: the only requirement is to keep the same unit (\si{\nano\metre} in this example) along the whole script. An additional optional parameter that defines the radiation intensity is described later in the advanced example in subsection \ref{subsec:sim_example2}.
\paragraph{Define a detector (line \num{17}):} The scatman module provides three detector types: \verb+MSFT+, \verb+Ideal+ and \verb+MCP+. \verb+MSFT+ is a \emph{virtual} detector, which directly yields the plain MSFT calculation, while the \verb+Ideal+ one attempts to model realistic photon statistics and noise augmentation. The \verb+Ideal+ detector is described as part of the advanced example in subsection \ref{subsec:sim_example2}. Finally, the PyScatman provides the \verb+MCP+ class, which aims at simulating a scattering detector based on a microchannel plate (MCP) \cite{Hamamatsu2019}, often used in CDI experiments \cite{Bostedt2010, rupp2017coherent, langbehn2018three, Rupp2020}. The \verb+MCP+ class is not described here as it is beyond the scope of this manuscript. However, a full description can be found in the \emph{Detectors} section in the official PyScatman documentation\footnote{\url{https://nux-group.gitlab.io/pyscatman/detectors.html}}.

In this fundamental example, the \verb+MSFT+ detector is used, which returns the exact MSFT calculation.

\begin{figure}
  \centering
  \includegraphics[width=\linewidth]{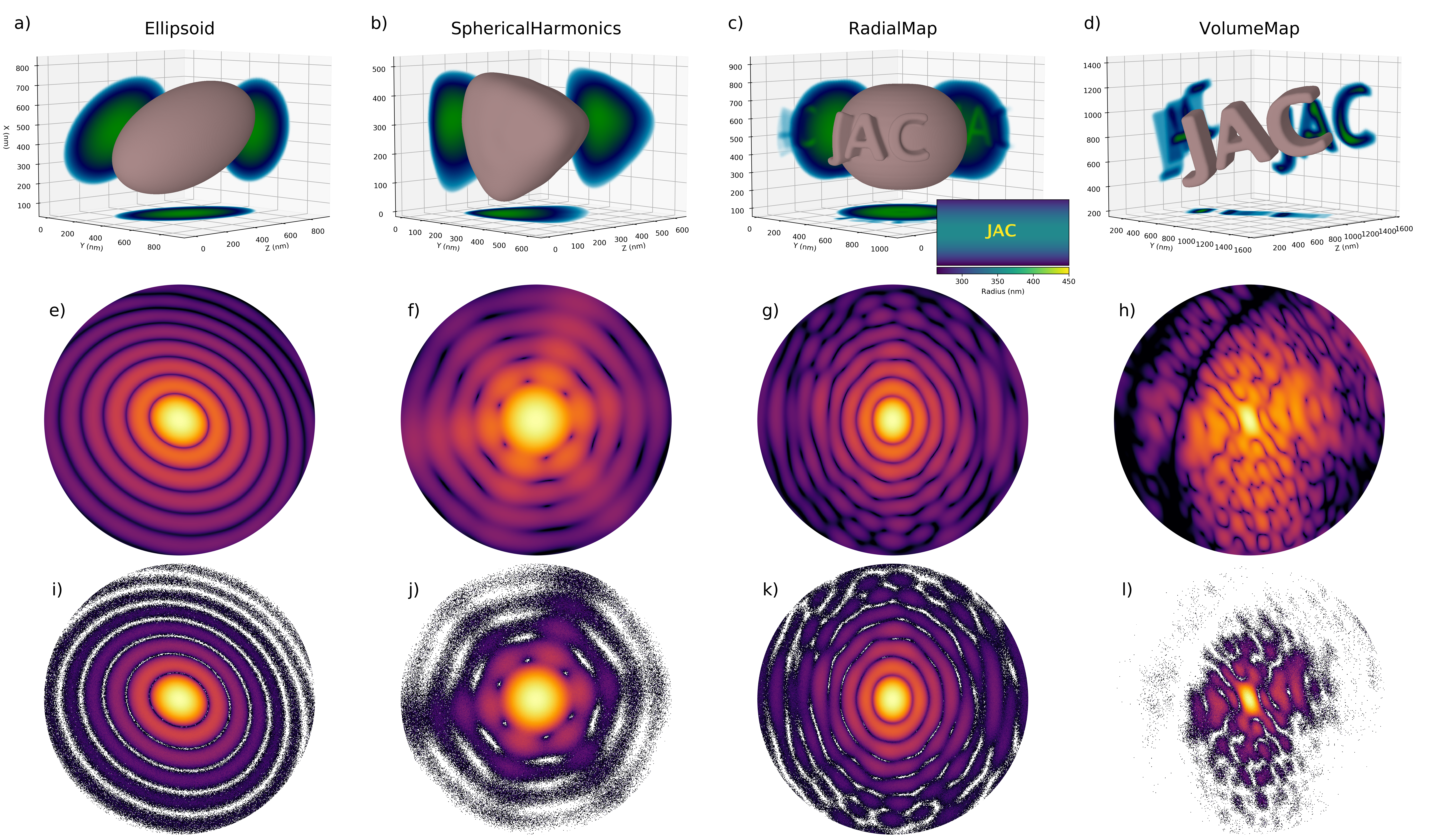}
  \caption{Rendering of the shapes and their respective simulated diffraction patterns using two different detectors. From a) to d), the 3D rendering of the shape objects, defined using the \emph{Ellipsoid} (a), \emph{SphericalHarmonics} (b), \emph{RadialMap} (c) and \emph{VolumeMap} (d) models, respectively. The \emph{RadialMap} example in c) has an inset showing the array that was used for creating the shape, where the radius information is color coded. The corresponding diffraction patterns of samples a) to d), computed by PyScatman via the \emph{MSFT} detector, are shown in e) to h). The third row, from i) to l), shows instead the equivalent simulations' results provided by the \emph{Ideal} detector. Here, the effects of photon statistics is clearly visible, along with the dependence on the value of the absorption coefficient. For example, the samples in b) and c) have an absorption coefficient $\beta=0.01$ and $\beta=0.05$ respectively, which reflect into a signal to noise ratio higher in k) than in j). Refer to the examples in the main text for further details.}
  \label{fig:examples}
\end{figure}

\paragraph{Define a shape (line \numrange{24}{33}):} PyScatman comes with several  pre-defined sample shape models, each with specific parameters that define their appearance\footnote{Please find a detailed explanation of every shape, along with images showing each axes, in the official documentation under \emph{The Shapes} at: \url{https://nux-group.gitlab.io/pyscatman/shapes.html}.}. The sample described in listing \ref{script:pyscatman1} is of \verb+Ellipsoid+ shape and is shown in Fig.  \ref{fig:examples} a). Note that the three axes of the ellipsoid are given in \si{\nano\metre} unit, as they must be consistent with the definition of the radiation wavelength set at line \num{7}.

\begin{figure}
  \centering
  \includegraphics[width=0.7\linewidth]{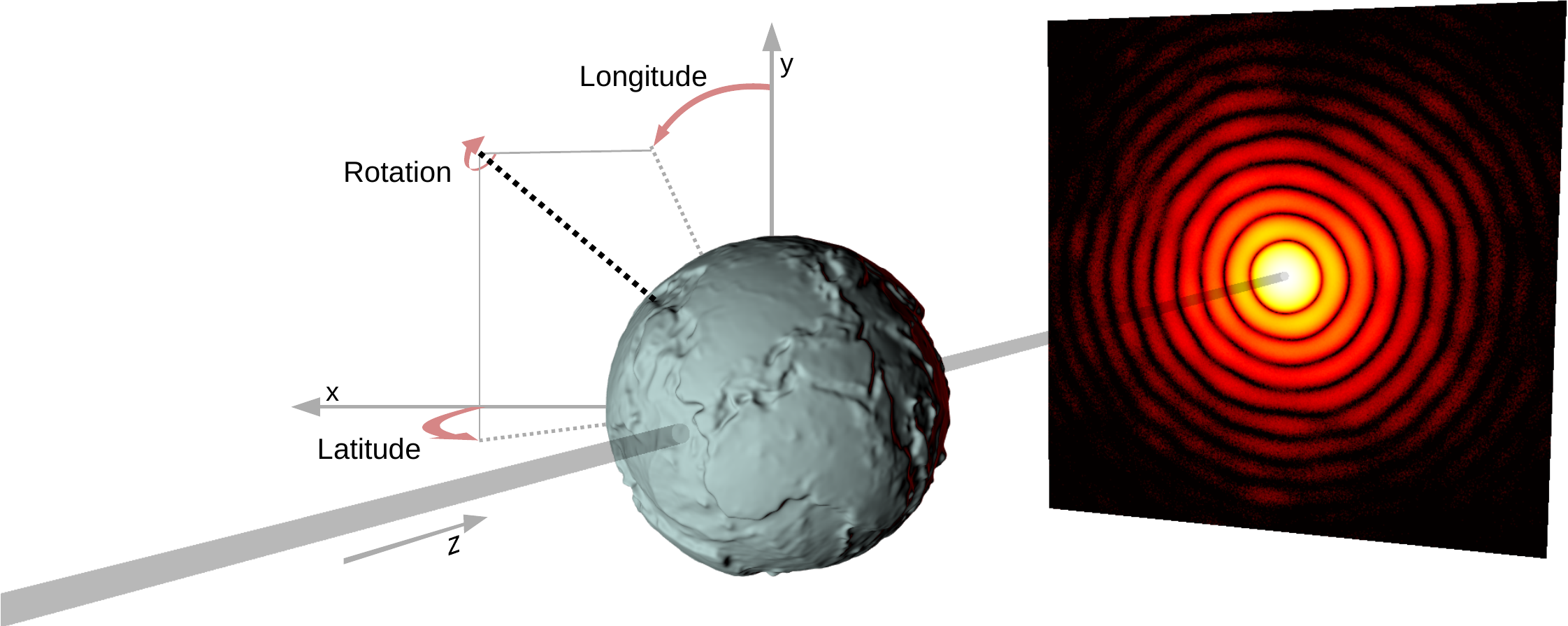}
  \caption{Schematic of the orientation in space for the \emph{latitude}, \emph{longitude}, and \emph{rotation} properties for the shapes in the PyScatman module. A shape model is oriented by setting the direction in space of its main axis. This direction is defined through \emph{latitude} and \emph{longitude} parameters, expressed on a reference system where $90^\circ$ latitude indicates a direction towards the incoming beam and, thus, $-90^\circ$ latitude towards the detector. The \emph{equator} of the reference system, at $0^\circ$ latitude, lies on the plane orthogonal to the beam. Once that the shape's main axis is oriented, a \emph{rotation} is applied to the sample along this axis. 
  Here, the simulated sample reflects Earth's \emph{elevation}, to highlight the large flexibility that the PyScatman module offers in terms of how to define the target shape. The sample was modeled though an adaptation of \emph{ETOPO1} data \cite{amante2009etopo1}, here in an exaggerated scale, which provides Earth's elevation as function of the Earth's coordinates, and simulated by the use of the \emph{RadialMap} shape model. Please see listing \ref{script:pyscatman2} for further details on how to define such a shape.}  
  \label{fig:shape_orientation}
\end{figure}

All shapes have a \verb+delta+, \verb+beta+, \verb+latitude+, \verb+longitude+, and \verb+rotation+ preference, which define their refractive index inside the sample and their orientation in space (see Fig. \ref{fig:shape_orientation} for a schematic on how the coordinates are defined). There, the \verb+latitude+ and \verb+longitude+ properties follow the standard convention used also for defining the coordinates on Earth, where the North-South axis is along the $z$ direction (solid gray line).

\paragraph{Calculate the MSFT (line \numrange{41}{42}):} After having defined a shape and a detector for an experiment, we can use the \verb+acquire+ method of the \emph{detector} class to calculate the MSFT-based diffraction image. In this example, the variable \verb+pattern_el+ is a Numpy array with dimensions \num{1024x1024}, as this was the resolution set at line \num{9}. The final calculation of the diffraction image is shown in Fig. \ref{fig:examples} e).

\subsection{A more advanced example}
\label{subsec:sim_example2}

One of the main advantages of the PyScatman module is the flexibility with which any arbitrary shape can be defined. In addition to the pre-defined shapes introduced in section \ref{subsec:sim_example1}, here we present three additional methods that PyScatman provides for defining an arbitrary shape: (i) \emph{SphericalHarmonics}, (ii) a \emph{RadialMap} and (iii) \emph{VolumeMap}. All three methods are described in listing \ref{script:pyscatman2}.

\begin{lstlisting}[language=Python, caption={A more advanced PyScatman example. Here, we calculate the MSFT diffraction images for various samples whose shapes are defined each in a different way.}, label=script:pyscatman2]
import scatman

"""
We re-use the experiment from listing 1. An optional parameter to define the photon density is added.
"""
scatman.set_experiment(
    wavelength=40,           # in nm
    angle=30,                # in degree
    resolution=512           # in pixels
    photon_density=1.e2      # in counts per squared nm
)

"""
We define an 'Ideal' detector.
See text in Section 5.2 for a full explanation.
"""
detector = scatman.Detectors.Ideal()

"""
Here, we define three additional shapes, using three
different methods.

Shape (1/3): Define an arbitrary shape using spherical
             harmonics coefficients. See Fig. 7b)
"""
shape_sh = scatman.Shapes.SphericalHarmonics(
    coefficients=[[0, 0, 200.0],  
                  [3, 2, 20.0],   
                  [6, 4, 4.0],
                  [9, 6, 2.0]],
    delta=0.01,             # refractive index
    beta=0.001,
    latitude=30,            # the orientation in space
    longitude=-40,          # see Fig. 6
    rotation=0
)

"""
Shape (2/3):  Define an arbitrary shape using a radial map.
              See Fig. 7c)
"""
shape_rm = scatman.Shapes.RadialMap(
    radii=radial_map_data,  
    delta=-0.01,            # refractive index
    beta=0.01,
    latitude=0,             # the orientation in space
    longitude=180,          # see Fig. 6
    rotation=30
)

"""
Shape (3/3): Define an arbitrary shape using a volume map.
             See Fig. 7 d)
"""
shape_vm = scatman.Shapes.VolumeMap(
    dx=6.5,
    data=volume_data,  
    delta=-0.001,           # refractive index
    beta=0.001,
    latitude=60,            # the orientation in space
    longitude=-70,          # see Fig. 6
    rotation=30
)

"""
Perform the MSFT calculation in parallel for all three shapes.
"""
patterns = detector.acquire(
    shape=[shape_sh, shape_rm, shape_vm]
)
\end{lstlisting}

\paragraph{Defining an \emph{Ideal} detector (line \numrange{1}{17}):} We import the PyScatman module and set up the same experiment as in listing \ref{script:pyscatman1}, with the addition of the optional parameter \verb+photon_density+ that plays a role in the later-defined \verb+Ideal+ detector. The idea behind the implementation of the \verb+Ideal+ detector is that even a perfect real-life detector is subjected to Poisson statistics of photons, which augments the recorded diffraction images. In order to model this effect, an estimate on the amount of scattered photons has to be calculated, and then used to add the proper Poisson noise to the simulated diffraction pattern. A description of how this data augmentation is implemented in PyScatman is given in the Supplemental Material.

\paragraph{Shape (1/3) via spherical harmonic coefficients (line \numrange{26}{36}):}

Any shape that is described by a radius as function of the azimuthal and polar angles, can be also defined using spherical harmonic coefficients. A notable example is the equipotential surface of the gravity potential of the Earth, which is termed the \emph{Geoid} and is defined using spherical harmonics \cite{Barthelmes2009}.

In general, the convention we use for the Laplace spherical harmonics ($Y_{\ell}^{m}$) are defined as:
\begin{equation}
  Y_\ell^m( \vartheta , \varphi ) = \sqrt{{\frac{(2\ell+1)}{4\pi}}{\frac{(\ell-m)!}{(\ell+m)!}}}  \, P_\ell^m ( \cos{\vartheta} ) \, e^{i m \varphi },
\end{equation}
where $m$ and $l$ are the order and degree of the harmonics, $\vartheta$ and $\varphi$ are the azimuthal and polar angles within the spherical coordinate system, and $P_\ell^m$ are the associated Legendre polynomials defined as:
\begin{equation}
  P_\ell ^{-m} = (-1)^m \frac{(\ell-m)!}{(\ell+m)!} P_\ell ^{m} \,.
\end{equation}

The PyScatman's \verb+SphericalHarmonics+ class expects a list of triplets, where the first value corresponds to degree $l$, the second value to the order $m$, and the third value to a scaling parameter with which $Y_\ell^m( \theta , \varphi )$ is multiplied. The final shape is then the sum of all triplets within the passed list.

The shape defined at line \numrange{26}{36} in listing \ref{script:pyscatman2} can be seen in Fig. \ref{fig:examples} b), along the calculated MSFT diffraction image for this shape in Fig. \ref{fig:examples} f).

\paragraph{Shape (2/3) via a radii map (line \numrange{42}{49}):}
A second method for defining an arbitrary shape within the PyScatman is to provide a two-dimensional array of any size that holds the length of the radii for all values of both angles $\theta$ and $\varphi$, which can be interpreted as the \emph{latitude} and \emph{longitude} coordinates. For example, when an array with size \SI{4x4}{\pixel} is passed, then these values define the radii of the shape at the $\theta$ values $\nicefrac{-\pi}{2}$, $\nicefrac{-\pi}{4}$, $0$, and $\nicefrac{\pi}{4}$, and for $\varphi$ at the values $0$, $\nicefrac{\pi}{2}$, $\pi$, and $\nicefrac{3 \pi}{2}$. These values are then linearly interpolated when the sample shape is rendered at the proper resolution, depending on the sample size and the \emph{experimental} conditions defined at line \numrange{6}{11}.

The shape of type \verb+RadialMap+ presented in the example, named \verb+shape_rm+, is produced by the 2D radii map \verb+radial_map_data+ (of size \num{1920x960}) given as argument in line \num{43}. The rendered shape can be found in Fig.  \ref{fig:examples} c), where an inset shows the used radii map. The MSFT calculation for this shape is shown in Fig.  \ref{fig:examples} g).

\paragraph{Shape (3/3) via a 3D volume map (line \numrange{55}{63}):}
The third method for defining an arbitrary shape is via a so-called volume map. The \verb+VolumeMap+ class of PyScatman requires a three-dimensional array of boolean type (\verb+volume_data+ at line 59, here of size \num{200x100x50}), which can have any size. The \verb+dx+ parameter, then, defines the linear size of a single volume unit of the 3D array \verb+volume_data+, and, as usual, must be expressed in the same length unit of the wavelength. 
For example, if a \SI{10x10x10}{\pixel} array with every value set as boolean \emph{True} is passed as \verb+data+ argument, and the \verb+dx+ argument is set to \num{2}, we end up with a cubic shape of size \SI{20x20x20}{\nano\metre}. If we want to scale up that cube by a factor of \num{2}, we can set the \verb+dx+ property to \num{4}, which results in a cube with doubled dimensions. 
The 3D rendering of the shape defined in this example is presented in Fig. \ref{fig:examples} d), along with its MSFT simulation in Fig.  \ref{fig:examples} h). PyScatman also provides the possibility to perform simulations of non-uniform samples via a completely arbitrary voxel representation of the refractive index . This feature is not presented here for sake of simplicity: for more information please visit the software documentation.

\paragraph{Obtaining the results (line \numrange{70}{72}):}
Finally, the simulation is performed for all three shapes. All shapes can be simulated through a single call to the \verb+Ideal+ detector's \verb+acquire+ method, passing them as a list. This possibility is implemented for allowing the PyScatman module to better exploit parallel computing hardware (and especially multiple GPUs) when large datasets have to be simulated (see Section \ref{sec:performance} for further details). The \verb+patterns+ array yields the simulations, formatted as a list of 2D arrays that contain the simulation result for the shapes \verb+shape_sh+, \verb+shape_rm+ and \verb+shape_vm+. These patterns are depicted in Fig.  \ref{fig:examples} j), \ref{fig:examples} k) and \ref{fig:examples} l) respectively, where the effects of photon statistics simulated by the \verb+Ideal+ detector are clearly visible.

\section{Performance considerations}
\label{sec:performance}

Our primary intention for the PyScatman module is to enable data analysis on diffraction patterns by forward-fitting the MSFT simulation with the experimental results, since \emph{classical} Fourier reconstruction via \emph{phase retrieval} methods is not possible for wide-angle scattering. The model fitting approach consists in \emph{guessing} a target's shape, simulating its diffraction pattern, comparing it with the desired experimental data, and then iteratively improving the guess until the MSFT simulation is sufficiently close to the experimental image.
Such an optimization scheme is computationally expensive in its own right. Therefore, it is of utmost importance to speed up the MSFT simulation as much as possible.

To this end, we provide in this section an overview on some benchmark results on CPU and GPU, based on the examples shown in listings \ref{script:pyscatman1} and \ref{script:pyscatman2}.

A dissection of the total computational cost of the MSFT routine reveals that time consumption of the simulation is mostly determined by the amount of Discrete Fourier Transforms (DFTs) (one for each slice), and the target's rendering process. Time complexity of a single DFT is given by $C_\text{DFT} \sim O[N_p^2\,\log(N_p^2)]$ \cite{cooley1965algorithm}, where $N_{p}$ is the resolution of the output image along a single axis. The complexity for MSFT algorithm scales linearly with the number of slices ($N_s$), so that the total time complexity of the MSFT's DFT part scales with $C_\text{MSFT, DFT} \sim  O[N_s \, N_p^2\,\log(N_p^2)]$.
Moreover, the time complexity of the rendering process ($C_\text{MSFT, Render}$) can be roughly estimated as $C_\text{MSFT, Render} \sim  O[N_s^3]$.

These considerations show that the resolution of the output image and the spatial extension of the sample, on which $N_p$ and $N_s$ respectively depend, are the determining factors for the running time of a PyScatman simulation.

When a single shape object is given to the detector's \emph{acquire} method, the PyScatman module carries out the MSFT simulation differently depending on the available hardware:
\begin{itemize}
  \item \textbf{CPU-only systems}: Slice rendering is sequential, where each slice is rendered using all CPU cores in parallel. After all $N_{s}$ slices are rendered, all CPUs perform the DFT calculations using the \emph{embarrassingly parallel} scheme.
  \item \textbf{Single NVIDIA\textsuperscript{\textregistered} GPU}: Each slice rendering and its respective DFT calculation are performed in parallel by the CUDA cores. Only one CPU is used for taking care of data preparation, inter-process communication and merging.
\end{itemize}
Therefore, if only a single shape is passed to the detector's \verb+acquire+ method, as it happens in Listing \ref{script:pyscatman1}, then, even in the case of a multi-GPU system, only one GPU is used, as the overhead caused by data transfers between the different GPUs' memory would prevent a performance scaling.

However, when multiple shapes are to be simulated, as the example presented in Listing \ref{script:pyscatman2}, multiple GPUs can speed up computation:
\begin{itemize}
  \item \textbf{CPU-only systems}: The multiple shapes are split evenly between all available CPU cores, where, subsequently, each core takes care of performing the shape rendering and the DFT calculations.
  \item \textbf{Single NVIDIA\textsuperscript{\textregistered} GPU}: Similarly to the \textbf{CPU-only} case, the multiple shapes are split evenly between all available CPU cores. Each CPU, then, properly set up the shape's data and submits the work to the GPU, where the CUDA cores calculate the rendering and the DFTs for all slices. 
  \item \textbf{Multiple NVIDIA\textsuperscript{\textregistered} GPUs}: Similar to the \textbf{Single GPU} case with the exception that the available CPUs are placed in groups where each group has an assigned GPU. For example, in an eight CPU cores / four GPUs system, two CPUs would share a single GPU and coordinate as in the \textbf{Single GPU} case.
\end{itemize}

Here, we present some benchmark results that we consider representative of \emph{real-life} situations. First, it is worth noting that the amount of computation, and thus the time to solution, depends on some factors (with most of them that can be deduced from Fig. \ref{fig:flowchart}):
\begin{itemize}
  \item \textbf{Simulation resolution}: the greater the number of pixels in the output image, the greater the computational cost.
  \item \textbf{Shape extension}: The greater the ratio between the sample size and the wavelength $\lambda$, the higher the amount of shape voxels to be rendered. Moreover, a greater scattering angle corresponds to a greater spatial resolution, such that the amount of pixels to be rendered increases accordingly to the maximum scattering angle $\theta_\text{max}$.
  \item \textbf{Shape complexity}: the function that defines the shape optical properties, $\delta(x,y,z)$ and $\beta(x,y,z)$ have a non-negligible computational cost, depending both on the shape type and the input data.
\end{itemize}

Among these three aspects, the contribution of the shape complexity to the total computing time is the less straightforward to evaluate in a systematic and quantitative manner, as it highly depends on the shape type and the values of its parameters. For example, the \verb+SphericalHarmonics+ complexity is particularly low when only few harmonics coefficients are provided as input: as the amount of harmonics coefficients increases, the data preparation step, which consists in the computation of the Spherical Harmonics transform, starts to take a relevant part in the computation time. The same happens, for example, for the \verb+VolumeMap+ object, for which the time dedicated to data transfer has an effect on the time to solution, depending on the size of the 3D array given as input. The authors are convinced that such a \emph{case-by-case} study goes beyond the scope of this manuscript, and encourage the reader to install the PyScatman module and test it for the cases of interest.

However, to give a rough idea about performance for different sample shapes, a first test is performed on the same four shape objects defined in the examples of the previous section, that are \verb+shape_el+ (\verb+Ellipsoid+),  \verb+shape_sh+ (\verb+SphericalHarmonics+), \verb+shape_rm+ (\verb+RadialMap+) and \verb+shape_vm+ (\verb+VolumeMap+), keeping the same experimental conditions and detector resolution. Here, the detector used is the \verb+MSFT+ one, yielding the diffraction patterns shown in Fig. \ref{fig:examples} e) to \ref{fig:examples} h).
The performance evaluation was accomplished on a workstation, equipped with an Intel\textsuperscript{\textregistered}  Core\textsuperscript{TM}  i9-9900K CPU accelerated by a GPU NVIDIA\textsuperscript{\textregistered} GeForce\textsuperscript{\textregistered} RTX 2080 Ti.

\begin{figure}
  \centering
  \includegraphics[width=\linewidth]{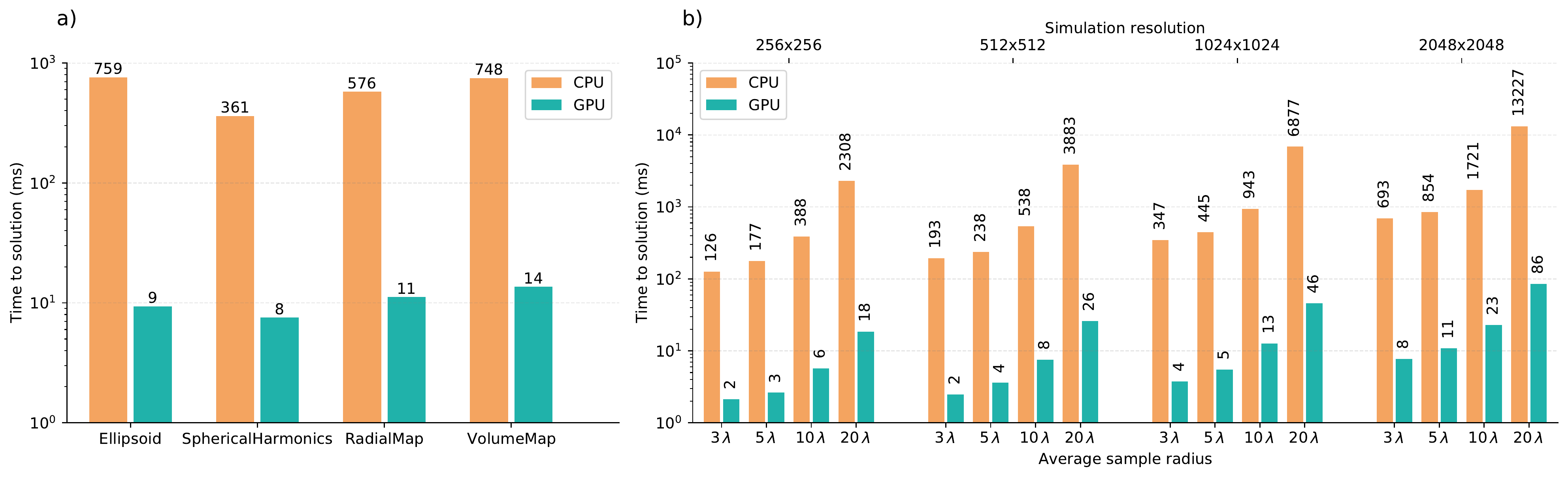}
  \caption{In a), timing results for the shapes presented as examples in Listing \ref{script:pyscatman1} and Listing \ref{script:pyscatman2}. The time to solution is shown in logarithmic scale versus the different shape types. In b), timing results for a shape defined through the \emph{SperhicalHarmonics} model are given. The harmonics coefficients are the same of the example presented in the main text, but the shape is scaled to get different average radii, indicated on the lower $x$ axis in units of the wavelength $\lambda$. Results are presented for four different simulation resolutions, from $256\times 256$ up to $2048 \times 2048$ pixel, labeled on the upper $x$ axis. }
  \label{fig:timing}
\end{figure}

The simulation time is shown in Fig. \ref{fig:timing} a). On the $x$ axis the four different shape models are labeled. The time to solution is on the $y$ axis, expressed in milliseconds on a logarithmic scale. The time shown is the execution time of a single call to the detector's \verb+acquire+ method, with a single shape object given as argument and averaged over 100 repetitions to rule out statistical fluctuations.
Two features are evident in the figure: first, the performance difference between the CPU time and the respective GPU one is around two orders of magnitude. Second, the time to solution depends on the shape. The first observation well underlines why the PyScatman implementation, when executed on a GPU, really enables a new kind of data analysis with the Scatman approach. The second feature, instead, is due to different shape sizes and complexities.

To quantitatively investigate the dependence of simulation time on the detector resolution and the sample's spatial extension, a second test is presented in Fig. \ref{fig:timing} b) . All the timing values in this test are based upon the same sample shape rendered in Fig. \ref{fig:examples} b), defined through the \verb+SphericalHarmonics+ model. Here, that shape is \emph{scaled} to match different average radii, that are $3\lambda$, $5\lambda$, $10\lambda$ and $20\lambda$, in order to get different sample spatial extensions without varying their \emph{complexity}. For each of them, two evaluations of the time to solution are performed, one running on the CPU and the other on the GPU. The whole operation is repeated for four different detector's resolutions, that are $256\times 256$, $512\times 512$, $1024\times 1024$ and $2048 \times 2048$ pixel. 
Here, again, the great advantage gained through the GPU computing arises. In particular, it is worth noting that the difference of around two orders of magnitude in the simulation time between GPU and CPU is  consistently present for all the different sample's sizes and the resolutions of the diffraction patterns, with the GPU that is still capable of performing more than ten simulations per second even in the worst, most complex case.

The presented timing results show the performance of the PyScatman module at the current stage. The software is, however, still under development, and better timing performances are expected in future software releases thanks to a better optimization of the GPU management.

\section{Summary}

In this paper we introduced the Scatman, an approximate method to simulate wide-angle diffraction patterns from coherent and monochromatic light based on the Multi-Slice Fourier Transform. The scientific impact of the method is already demonstrated by previous publications that made use of the Scatman, while it was under development, to retrieve three dimensional morphological information on silver nanocrystals \cite{barke20153d} and helium nanodroplets \cite{langbehn2018three} from single wide-angle diffraction images.

The need for an approximate simulation tool arises from the severe limitations of the available exact methods: Mie calculations, which are fast but can be used only for highly symmetrical samples, and Finite Difference Time Domain or Discrete Dipole Approximation simulations, which are computationally heavy.  The Scatman was conceived to be both generic, i.e. capable of simulating the scattering from any kind of sample, and sufficiently fast, enabling the retrieval of the sample morphology by fitting the experimental diffraction patterns via a \emph{model fitting} approach.

The mathematical foundations of the method were presented, highlighting the main approximations that make the Scatman results deviate from the exact ones. The effects of these approximations as function of the input parameters were investigated, by comparing the Scatman simulations and the exact Mie calculations for a spherical sample. Within given bounds on the optical properties of the sample and its spatial extension, the Scatman results proved to be in quantitative agreement with exact calculations.

We presented our reference implementation of the Scatman, called PyScatman, that is released as Open Source software with this manuscript and is freely available online. PyScatman, implemented as a Python module, provides an easy interface to the user and a set of additional functionalities useful for data analysis. PyScatman is entirely written in C++ and makes use of state of the art programming techniques to take full advantage from the most recent computing hardware, including GPU accelerators. The computational performance of PyScatman was briefly presented, demonstrating the possibility to perform wide-angle scattering simulations in few to few tens of milliseconds on consumer-level computing hardware. These computing times well suit the extensive use of PyScatman in the analysis of experimental data via forward-fitting procedures, thus opening new perspectives for Coherent Diffraction Imaging in wide-angle scattering conditions.

The Scatman method described here is a stable and tested snapshot of its current development. Further enhancements are under study, focusing on both the physics aspect and the software implementation. In terms of software, the inclusion of additional, more sophisticated and ductile shape models is planned, along with a more efficient management of computing resources. On the physics side, the partial inclusion of secondary effects like multiple scattering, refraction and reflection is under study, extending the range of applicability of the approach to samples whose refractive index varies more strongly from unity.

The Scatman method and its software implementation aims at being a reference tool for the Coherent Diffraction Imaging community, and can be of great interest for other scientific fields where elastic scattering of coherent radiation plays a role, like the recently growing Electron Diffraction Imaging techniques. Moreover, the high-performance software implementation, PyScatman, well suits the increasing interest in big-data analysis and Artificial Intelligence, and its combination with AI techniques is already in a prototyping phase.

\paragraph{Acknowledgments}

This project was funded by the Swiss National Science Foundation via the NCCR MUST.
Further funding is acknowledged from: DFG grants MO 719/13-1 and MO 719/14-2; the SNF grant No. 200021E\_193642 ; the Deutsche Forschungsgemeinschaft via SFB 652 and SFB 1477;  Heisenberg-Grant (ID: 436382461) via SPP1840 (ID: 281272685); the Bundesministerium für Bildung und Forschung (BMBF, ID: 05K16HRB); the European Social Fund and the Ministry of Education, Science and Culture of Mecklenburg-Vorpommern, Germany, via project NEISS (ID: ESF/14-BM-A55-0007/19).

\bibliographystyle{unsrt}
\bibliography{main}

\begin{thebibliography}{10}

\bibitem{chapman2010coherent}
Henry~N Chapman and Keith~A Nugent.
\newblock Coherent lensless x-ray imaging.
\newblock {\em Nature photonics}, 4(12):833--839, 2010.

\bibitem{miao2011coherent}
Jianwei Miao, Richard~L Sandberg, and Changyong Song.
\newblock Coherent x-ray diffraction imaging.
\newblock {\em IEEE Journal of selected topics in quantum electronics},
  18(1):399--410, 2011.

\bibitem{seibert2011single}
M~Marvin Seibert, Tomas Ekeberg, Filipe~RNC Maia, Martin Svenda, Jakob
  Andreasson, Olof J{\"o}nsson, Du{\v{s}}ko Odi{\'c}, Bianca Iwan, Andrea
  Rocker, Daniel Westphal, et~al.
\newblock Single mimivirus particles intercepted and imaged with an x-ray
  laser.
\newblock {\em Nature}, 470(7332):78--81, 2011.

\bibitem{guinier1955small}
Andr{\'e} Guinier, G{\'e}rard Fournet, and Kenneth~L Yudowitch.
\newblock Small-angle scattering of x-rays.
\newblock 1955.

\bibitem{born1926quantenmechanik}
Max Born.
\newblock Quantenmechanik der sto{\ss}vorg{\"a}nge.
\newblock {\em Zeitschrift f{\"u}r Physik}, 38(11-12):803--827, 1926.

\bibitem{miao1999extending}
Jianwei Miao, Pambos Charalambous, Janos Kirz, and David Sayre.
\newblock Extending the methodology of x-ray crystallography to allow imaging
  of micrometre-sized non-crystalline specimens.
\newblock {\em Nature}, 400(6742):342--344, 1999.

\bibitem{fienup1982phase}
James~R Fienup.
\newblock Phase retrieval algorithms: a comparison.
\newblock {\em Applied optics}, 21(15):2758--2769, 1982.

\bibitem{marchesini2007invited}
Stefano Marchesini.
\newblock Invited article: A unified evaluation of iterative projection
  algorithms for phase retrieval.
\newblock {\em Review of scientific instruments}, 78(1):011301, 2007.

\bibitem{loh2012fractal}
ND~Loh, Christina~Y Hampton, Andrew~V Martin, Dmitri Starodub, Raymond~G
  Sierra, Anton Barty, Andrew Aquila, Joachim Schulz, Lukas Lomb, Jan
  Steinbrener, et~al.
\newblock Fractal morphology, imaging and mass spectrometry of single aerosol
  particles in flight.
\newblock {\em Nature}, 486(7404):513--517, 2012.

\bibitem{pedersoli2013mesoscale}
E~Pedersoli, ND~Loh, F~Capotondi, CY~Hampton, RG~Sierra, D~Starodub, C~Bostedt,
  J~Bozek, AJ~Nelson, M~Aslam, et~al.
\newblock Mesoscale morphology of airborne core--shell nanoparticle clusters:
  X-ray laser coherent diffraction imaging.
\newblock {\em Journal of Physics B: Atomic, Molecular and Optical Physics},
  46(16):164033, 2013.

\bibitem{miao2006three}
Jianwei Miao, Chien-Chun Chen, Changyong Song, Yoshinori Nishino, Yoshiki
  Kohmura, Tetsuya Ishikawa, Damien Ramunno-Johnson, Ting-Kuo Lee, and
  Subhash~H Risbud.
\newblock Three-dimensional gan- ga 2 o 3 core shell structure revealed by
  x-ray diffraction microscopy.
\newblock {\em Physical review letters}, 97(21):215503, 2006.

\bibitem{jiang2010quantitative}
Huaidong Jiang, Changyong Song, Chien-Chun Chen, Rui Xu, Kevin~S Raines,
  Benjamin~P Fahimian, Chien-Hung Lu, Ting-Kuo Lee, Akio Nakashima, Jun Urano,
  et~al.
\newblock Quantitative 3d imaging of whole, unstained cells by using x-ray
  diffraction microscopy.
\newblock {\em Proceedings of the National Academy of Sciences},
  107(25):11234--11239, 2010.

\bibitem{lundholm2018considerations}
Ida~V Lundholm, Jonas~A Sellberg, Tomas Ekeberg, Max~F Hantke, Kenta Okamoto,
  Gijs van~der Schot, Jakob Andreasson, Anton Barty, Johan Bielecki, Petr
  Bruza, et~al.
\newblock Considerations for three-dimensional image reconstruction from
  experimental data in coherent diffractive imaging.
\newblock {\em IUCrJ}, 5(5):531--541, 2018.

\bibitem{loh2010cryptotomography}
ND~Loh, Michael~J Bogan, Veit Elser, Anton Barty, S{\'e}bastien Boutet,
  Sa{\v{s}}a Bajt, Janos Hajdu, Tomas Ekeberg, Filipe~RNC Maia, Joachim Schulz,
  et~al.
\newblock Cryptotomography: reconstructing 3d fourier intensities from randomly
  oriented single-shot diffraction patterns.
\newblock {\em Physical review letters}, 104(22):225501, 2010.

\bibitem{loh2009reconstruction}
Ne-Te~Duane Loh and Veit Elser.
\newblock Reconstruction algorithm for single-particle diffraction imaging
  experiments.
\newblock {\em Physical Review E}, 80(2):026705, 2009.

\bibitem{ekeberg2015three}
Tomas Ekeberg, Martin Svenda, Chantal Abergel, Filipe~RNC Maia, Virginie
  Seltzer, Jean-Michel Claverie, Max Hantke, Olof J{\"o}nsson, Carl Nettelblad,
  Gijs Van Der~Schot, et~al.
\newblock Three-dimensional reconstruction of the giant mimivirus particle with
  an x-ray free-electron laser.
\newblock {\em Physical review letters}, 114(9):098102, 2015.

\bibitem{feldhaus2005x}
Josef Feldhaus, John Arthur, and JB~Hastings.
\newblock X-ray free-electron lasers.
\newblock {\em Journal of Physics B: Atomic, molecular and optical physics},
  38(9):S799, 2005.

\bibitem{harmand2013achieving}
M~Harmand, R~Coffee, Mina~R Bionta, Matthieu Chollet, D~French, D~Zhu,
  DM~Fritz, HT~Lemke, N~Medvedev, B~Ziaja, et~al.
\newblock Achieving few-femtosecond time-sorting at hard x-ray free-electron
  lasers.
\newblock {\em Nature Photonics}, 7(3):215--218, 2013.

\bibitem{barty2013molecular}
Anton Barty, Jochen K{\"u}pper, and Henry~N Chapman.
\newblock Molecular imaging using x-ray free-electron lasers.
\newblock {\em Annual review of physical chemistry}, 64, 2013.

\bibitem{chapman2006femtosecond}
Henry~N Chapman, Anton Barty, Michael~J Bogan, S{\'e}bastien Boutet, Matthias
  Frank, Stefan~P Hau-Riege, Stefano Marchesini, Bruce~W Woods, Sa{\v{s}}a
  Bajt, W~Henry Benner, et~al.
\newblock Femtosecond diffractive imaging with a soft-x-ray free-electron
  laser.
\newblock {\em Nature Physics}, 2(12):839--843, 2006.

\bibitem{Chapman2014}
Henry~N Chapman, Carl Caleman, and Nicusor Timneanu.
\newblock Diffraction before destruction.
\newblock {\em Philosophical Transactions of the Royal Society B: Biological
  Sciences}, 369(1647):20130313, 2014.

\bibitem{xu2014single}
Rui Xu, Huaidong Jiang, Changyong Song, Jose~A Rodriguez, Zhifeng Huang,
  Chien-Chun Chen, Daewoong Nam, Jaehyun Park, Marcus Gallagher-Jones, Sangsoo
  Kim, et~al.
\newblock Single-shot three-dimensional structure determination of nanocrystals
  with femtosecond x-ray free-electron laser pulses.
\newblock {\em Nature communications}, 5(1):1--9, 2014.

\bibitem{barke20153d}
Ingo Barke, Hannes Hartmann, Daniela Rupp, Leonie Fl{\"u}ckiger, Mario Sauppe,
  Marcus Adolph, Sebastian Schorb, Christoph Bostedt, Rolf Treusch, Christian
  Peltz, et~al.
\newblock The 3d-architecture of individual free silver nanoparticles captured
  by x-ray scattering.
\newblock {\em Nature communications}, 6(1):1--7, 2015.

\bibitem{raines2010three}
Kevin~S Raines, Sara Salha, Richard~L Sandberg, Huaidong Jiang, Jose~A
  Rodr{\'\i}guez, Benjamin~P Fahimian, Henry~C Kapteyn, Jincheng Du, and
  Jianwei Miao.
\newblock Three-dimensional structure determination from a single view.
\newblock {\em Nature}, 463(7278):214--217, 2010.

\bibitem{wang2011non}
Ge~Wang, Hengyong Yu, Wenxiang Cong, and Alexander Katsevich.
\newblock Non-uniqueness and instability of ‘ankylography’.
\newblock {\em Nature}, 480(7375):E2--E3, 2011.

\bibitem{miao2011potential}
Jianwei Miao, Chien-Chun Chen, Yu~Mao, Leigh~S Martin, and Henry~C Kapteyn.
\newblock Potential and challenge of ankylography.
\newblock {\em arXiv preprint arXiv:1112.4459}, 2011.

\bibitem{hahn2009light}
David~W Hahn.
\newblock Light scattering theory.
\newblock {\em Department of Mechanical and Aerospace Engineering, University
  of Florida}, 2009.

\bibitem{bohren2008absorption}
Craig~F Bohren and Donald~R Huffman.
\newblock {\em Absorption and scattering of light by small particles}.
\newblock John Wiley \& Sons, 2008.

\bibitem{aden1951scattering}
Arthur~L Aden and Milton Kerker.
\newblock Scattering of electromagnetic waves from two concentric spheres.
\newblock {\em Journal of Applied Physics}, 22(10):1242--1246, 1951.

\bibitem{taflove1980application}
Allen Taflove.
\newblock Application of the finite-difference time-domain method to sinusoidal
  steady-state electromagnetic-penetration problems.
\newblock {\em IEEE Transactions on electromagnetic compatibility},
  (3):191--202, 1980.

\bibitem{varin2012attosecond}
Charles Varin, Christian Peltz, Thomas Brabec, and Thomas Fennel.
\newblock Attosecond plasma wave dynamics in laser-driven cluster nanoplasmas.
\newblock {\em Physical review letters}, 108(17):175007, 2012.

\bibitem{purcell1973scattering}
Edward~M Purcell and Carlton~R Pennypacker.
\newblock Scattering and absorption of light by nonspherical dielectric grains.
\newblock {\em The Astrophysical Journal}, 186:705--714, 1973.

\bibitem{sander2015influence}
Katharina Sander, Christian Peltz, Charles Varin, Stefan Scheel, Thomas Brabec,
  and Thomas Fennel.
\newblock Influence of wavelength and pulse duration on single-shot x-ray
  diffraction patterns from nonspherical nanoparticles.
\newblock {\em Journal of Physics B: Atomic, Molecular and Optical Physics},
  48(20):204004, 2015.

\bibitem{rupp2017coherent}
Daniela Rupp, Nils Monserud, Bruno Langbehn, Mario Sauppe, Julian Zimmermann,
  Yevheniy Ovcharenko, Thomas M{\"o}ller, Fabio Frassetto, Luca Poletto, Andrea
  Trabattoni, et~al.
\newblock Coherent diffractive imaging of single helium nanodroplets with a
  high harmonic generation source.
\newblock {\em Nature communications}, 8(1):1--7, 2017.

\bibitem{langbehn2018three}
Bruno Langbehn, Katharina Sander, Yevheniy Ovcharenko, Christian Peltz, Andrew
  Clark, Marcello Coreno, Riccardo Cucini, Marcel Drabbels, Paola Finetti,
  Michele Di~Fraia, et~al.
\newblock Three-dimensional shapes of spinning helium nanodroplets.
\newblock {\em Physical review letters}, 121(25):255301, 2018.

\bibitem{zimmermann2019deep}
Julian Zimmermann, Bruno Langbehn, Riccardo Cucini, Michele Di~Fraia, Paola
  Finetti, Aaron~C LaForge, Toshiyuki Nishiyama, Yevheniy Ovcharenko, Paolo
  Piseri, Oksana Plekan, et~al.
\newblock Deep neural networks for classifying complex features in diffraction
  images.
\newblock {\em Physical Review E}, 99(6):063309, 2019.

\bibitem{cowley1957scattering}
John~M Cowley and A~F\_ Moodie.
\newblock The scattering of electrons by atoms and crystals. i. a new
  theoretical approach.
\newblock {\em Acta Crystallographica}, 10(10):609--619, 1957.

\bibitem{self1983practical}
P~G\_ Self, MA~O'keefe, P\_R Buseck, and AEC Spargo.
\newblock Practical computation of amplitudes and phases in electron
  diffraction.
\newblock {\em Ultramicroscopy}, 11(1):35--52, 1983.

\bibitem{reinhard1997size}
D~Reinhard, BD~Hall, D~Ugarte, and R~Monot.
\newblock Size-independent fcc-to-icosahedral structural transition in
  unsupported silver clusters: An electron diffraction study of clusters
  produced by inert-gas aggregation.
\newblock {\em Physical Review B}, 55(12):7868, 1997.

\bibitem{hare1994near}
AR~Hare and GR~Morrison.
\newblock Near-field soft x-ray diffraction modelled by the multislice method.
\newblock {\em Journal of Modern Optics}, 41(1):31--48, 1994.

\bibitem{lambert1760photometria}
Johann~Heinrich Lambert.
\newblock {\em Photometria sive de mensura et gradibus luminis, colorum et
  umbrae}.
\newblock Klett, 1760.

\bibitem{beer1852bestimmung}
August Beer.
\newblock Bestimmung der absorption des rothen lichts in farbigen
  flüssigkeiten.
\newblock {\em Ann. Physik}, 162:78--88, 1852.

\bibitem{paganin2006coherent}
David Paganin et~al.
\newblock {\em Coherent X-ray optics}.
\newblock Number~6. Oxford University Press on Demand, 2006.

\bibitem{gaffney2007imaging}
KJ~Gaffney and Henry~N Chapman.
\newblock Imaging atomic structure and dynamics with ultrafast x-ray
  scattering.
\newblock {\em Science}, 316(5830):1444--1448, 2007.

\bibitem{cooley1965algorithm}
James~W Cooley and John~W Tukey.
\newblock An algorithm for the machine calculation of complex fourier series.
\newblock {\em Mathematics of computation}, 19(90):297--301, 1965.

\bibitem{Mie1908}
Gustav Mie.
\newblock {Beitr{\"{a}}ge zur Optik tr{\"{u}}ber Medien, speziell kolloidaler
  Metall{\"{o}}sungen}.
\newblock {\em Ann. Phys.}, 330(3):377--445, jan 1908.

\bibitem{Bohren1998}
Craig~F. Bohren and Donald~R. Huffman.
\newblock {\em {Absorption and Scattering of Light by Small Particles}}.
\newblock Wiley, apr 1998.

\bibitem{Rupp2014}
Daniela Rupp, Marcus Adolph, Leonie Fl{\"{u}}ckiger, Tais Gorkhover,
  Jan~Philippe M{\"{u}}ller, Maria M{\"{u}}ller, Mario Sauppe, David Wolter,
  Sebastian Schorb, Rolf Treusch, Christoph Bostedt, and Thomas M{\"{o}}ller.
\newblock {Generation and structure of extremely large clusters in pulsed
  jets}.
\newblock {\em J. Chem. Phys.}, 141(4):044306, jul 2014.

\bibitem{c++2017iso}
C++~Standards Committee et~al.
\newblock Iso international standard iso/iec 14882: 2017, programming language
  c++.
\newblock Technical report, Tech. rep. http://www. open-std.
  org/jtc1/sc22/wg21. Geneva, Switzerland~…, 2017.

\bibitem{jakob2019pybind11}
Wenzel Jakob, Jason Rhinelander, and Dean Moldovan.
\newblock pybind11--seamless operability between c++ 11 and python, 2017.
\newblock {\em URL https://github. com/pybind/pybind11}, 2019.

\bibitem{Nickolls2008}
John Nickolls, Ian Buck, Michael Garland, and Kevin Skadron.
\newblock {Scalable parallel programming with CUDA}.
\newblock {\em Queue}, 6(2):40--53, mar 2008.

\bibitem{Hamamatsu2019}
{Hamamatsu Photonics K.K.}
\newblock {MCP (Microchannel plate) and MCP assembly}.
\newblock Technical report, 2019.

\bibitem{Bostedt2010}
C.~Bostedt, M.~Adolph, E.~Eremina, M.~Hoener, D.~Rupp, S.~Schorb, H.~Thomas,
  A.~R.~B. {De Castro}, and T.~M{\"{o}}ller.
\newblock {Clusters in intense FLASH pulses: Ultrafast ionization dynamics and
  electron emission studied with spectroscopic and scattering techniques}.
\newblock {\em Journal of Physics B: Atomic, Molecular and Optical Physics},
  43(19):194011, oct 2010.

\bibitem{Rupp2020}
Daniela Rupp, Leonie Fl{\"{u}}ckiger, Marcus Adolph, Alessandro Colombo, Tais
  Gorkhover, Marion Harmand, Maria Krikunova, Jan~Philippe M{\"{u}}ller, Tim
  Oelze, Yevheniy Ovcharenko, Maria Richter, Mario Sauppe, Sebastian Schorb,
  Rolf Treusch, David Wolter, Christoph Bostedt, and Thomas M{\"{o}}ller.
\newblock {Imaging plasma formation in isolated nanoparticles with ultrafast
  resonant scattering}.
\newblock {\em Structural Dynamics}, 7(3):34303, may 2020.

\bibitem{amante2009etopo1}
Christopher Amante and Barry~W Eakins.
\newblock Etopo1 arc-minute global relief model: procedures, data sources and
  analysis.
\newblock 2009.

\bibitem{Barthelmes2009}
Franz Barthelmes.
\newblock {Definition of functionals of the geopotential and their calculation
  from spherical harmonic models : theory and formulas used by the calculation
  service of the International Centre for Global Earth Models (ICGEM)}.
\newblock Technical report, Deutsches GeoForschungsZentrum, Potsdam, sep 2009.

\end{thebibliography}


\end{document}